\documentclass[useAMS,usenatbib]{mn2e}

\usepackage{graphicx}
\usepackage{txfonts}
\usepackage[normalem]{ulem}
\usepackage{bm}

\usepackage{lscape}

\newcommand{\sax}{{\textit{Beppo\-SAX}}}
\newcommand{\integral}{{\textit{INTEGRAL}}}
\newcommand{\msun}{{\rm M}_{\sun}}
\newcommand{\xte}{{\textit{RXTE}}}
\def\xtee{{\it RXTE }}
\def\gro{{\it CGRO}}

\def\ginga{{\it Ginga}}

\def\swift{{\it Swift}}
\def\swifte{{\it Swift }}
\def\suzaku{{\it Suzaku}}

\def\asca{{\it ASCA}}

\def\xmm{{\it XMM-Newton}}
\def\xmme{{\it XMM-Newton }}
\def\chandra{{\it Chandra}}
\def\chandrae{{\it Chandra }}
\def\granat{{\it GRANAT}}

\def\rosat{{\it ROSAT}}

\def\hst{{\it HST}}

\title[Extreme flux states of NGC 4151]
{Extreme flux states of NGC 4151 observed with \integral\thanks{Based on observations with \integral, an ESA project with instruments and 
science data centre funded by ESA member states (especially the PI countries: 
Denmark, France, Germany, Italy, Switzerland and Spain), the Czech Republic 
and Poland and with participation of Russia and the US.}}

\author[P. Lubi\'nski et al.]{P. Lubi\'nski$^{1,2}$\thanks{E-mail: lubinski@ncac.torun.pl (PL), aaz@camk.edu.pl (AAZ)}, 
A. A. Zdziarski$^{3}$\footnotemark[2], 
R. Walter$^{2,4}$,
S. Paltani$^{2,4}$, 
V. Beckmann$^{5}$, 
S. Soldi$^{6}$,
\newauthor
C. Ferrigno$^{2}$
and T. J.-L. Courvoisier$^{2,4}$\\
$^{1}$Centrum Astronomiczne im. M. Kopernika, Rabia\'nska 8, PL-87-100 Toru\'n, 
      Poland\\
$^{2}$ISDC Data Centre for Astrophysics, Chemin d'Ecogia 16, CH-1290 Versoix, 
      Switzerland\\
$^3$Centrum Astronomiczne im. M. Kopernika, Bartycka 18, PL-00-716 Warszawa, 
      Poland\\
$^{4}$Observatoire de Gen\`eve, Universit\'e de Gen\`eve, Chemin des Maillettes 
      51, CH-1290 Sauverny, Switzerland\\
$^{5}$APC, UMR 7164, Universit\'e Paris 7 Denis Diderot, 10 rue Alice Domon et 
L\'eonie Duquet, F-75025 Paris Cedex 13, France\\      
$^{6}$Laboratoire AIM - CNRS - CEA/DSM - Universit\'e Paris Diderot (UMR 7158), 
CEA Saclay, DSM/IRFU/SAp, F-91191 Gif-sur-Yvette, France\\
}
      
\begin{document}

\date{Accepted 2010 June 25.  Received 2010 June 25; in original form 2010 May 4}

\pagerange{\pageref{firstpage}--\pageref{lastpage}} \pubyear{2010}

\maketitle

\label{firstpage}

\begin{abstract}
We present a comprehensive spectral analysis of all \integral\/ data obtained 
so far for the X-ray--bright Seyfert galaxy NGC 4151. We also use all 
contemporaneous data from \xte, \xmm, \swift\/ and \suzaku. We find a linear 
correlation between the medium and hard-energy X-ray fluxes measured by 
\integral, which indicates an almost constant spectral index over six years. 
The majority of \integral\/ observations were made when the source was either 
at a very bright or very dim hard--X-ray state. We find that thermal 
Comptonization models applied to the bright state yields the plasma 
temperature of $\simeq 50$--70 keV and its optical depth of $\simeq 1.3$--2.6, 
depending on the assumed source geometry. For the dim state, these parameters 
are in the ranges of $\simeq 180$--230 keV and $\simeq 0.3$--0.7, respectively. 
The Compton parameter is $y\simeq 1$ for all the spectra, indicating a stable 
geometry. Using this result, we can determine the reflection effective solid 
angles associated with the close and distant reprocessing media as $\simeq 
0.3\times 2\upi$ and $0.2 \times 2\upi$, respectively. The plasma energy 
balance, the weak disc reflection and a comparison of the UV fluxes 
illuminating the plasma to the observed ones are all consistent with an inner 
hot accretion surrounded by an outer cold disc. The disc truncation radius can 
be determined from an approximate equipartition between the observed UV and 
X-ray emission, and from the fitted disc blackbody model, as $\sim 15$ 
gravitational radii. Alternatively, our results can be explained by a mildly 
relativistic coronal outflow. 
\end{abstract}

\section{Introduction}
\label{intro}

Continuum properties of the hard X-ray and soft $\gamma$-ray emission from radio-quiet Seyfert galaxies nuclei are relatively well known thanks to the many satellites operating during the last decades. The spectra are commonly approximated by a phenomenological e-folded power-law model, $F(E)\propto E^{1-\Gamma} \exp(-E/E_{\rm c})$ (where $\Gamma$ is the photon index and $E_{\rm c}$ is the e-folding, or cut-off, energy), accompanied by a Compton reflection component. A recent study based on a local ($z < 0.1$) sample of 105 Seyfert galaxies observed with \sax\/ presents the average parameters of this model for Seyfert 1s and 2s \citep{Dadina2008}. The mean $\Gamma$ is $\approx 1.9$ (1.8) for Seyfert 1 (2) nuclei, $E_{\rm c}$ is $\simeq 230$ (380) keV, and the average relative strength of reflection, $R\equiv \Omega/2\upi$ (where $\Omega$ is the effective solid angle subtended by the reflector), is $\simeq 1.2$ (0.9). See \citet{Beckmann2009} for the similar \integral\/ results.

This phenomenological model is too crude for studying of the physics of the central engine of Seyferts. There is a general consensus that their X$\gamma$ photons are from (predominantly thermal) Comptonization by hot electrons of some seed soft photons. The seed photons may come from an optically-thick accretion disc or clouds in the vicinity of the hot plasma, or may be internally produced by the synchrotron process, e.g., \citet{Xie2010}. For example, non-simultaneous data for the Seyfert NGC 5548 from \rosat, \ginga\/ and \gro/OSSE have yielded the plasma temperature, $kT_{\rm e}\simeq 55$ keV, and the Thomson optical depth, $\tau\simeq 2$ \citep{Magdziarz1998}. The average OSSE spectra of 11 Sy 1s and 8 Sy 2s have yielded similar parameters, $kT_{\rm e}\simeq 70$--80 keV, $\tau\simeq 1.7$ \citep*{Zdziarski2000}. On the other hand, \citet{Petrucci2000} obtained a much higher $kT_{\rm e}\simeq 250$ keV (and $\tau\simeq 0.2$--0.4) for NGC 5548 using \sax\/ data and a different anisotropic Comptonization model in the slab geometry. Then, \citet{Petrucci2001a}, hereafter P01, found, with the same model, a similarly high values of $kT_{\rm e}\simeq 170$--320 keV (and $\tau\simeq 0.05$--0.20) for 6 Sy 1s observed by \sax. We note, however, that such high $kT_{\rm e}$ yield spectra being well above the OSSE fluxes at $\ga 100$ keV for either NGC 5548, NGC 4151 or the average Seyferts.   

The nearby, $z = 0.0033$, Seyfert 1.5 galaxy NGC 4151 is the second (after Cen 
A) brightest persistent AGN in the 20--100 keV band. Its electromagnetic 
spectrum has been extensively studied, with its properties in the radio, 
infrared, optical, ultraviolet and X$\gamma$ bands well established. Its 
black-hole mass has been estimated based on reverberation as 
$(4.6^{+0.6}_{-0.5}) \times 10^{7}\msun$ \citep{Bentz2006}. Although this estimate is subject to a systematic uncertainty by a factor of $\sim$3--4, it is consistent with the estimate based on scaling of the power spectrum to Cyg X-1 of \citet{Czerny2001}. The overall picture of the region surrounding the nucleus, based on {\it HST}/STIS, \chandrae and \xmme data, is that the complex 
absorption observed for NGC 4151 is due to massive outflows, presumably disc 
winds, forming several regions characterized by a range of column densities, 
$N_{\rm H}$, and ionization levels \citep{Schurch2004,Kraemer2006}. This 
explains earlier results from \rosat, \asca, \ginga\/ and \xte\/ satellites, 
with the medium-energy X-ray spectra fitted only after applying a complex 
absorber model, consisting of several components fully and partially covering 
the central source (e.g., \citealt*{Zdziarski1996a,Zdziarski2002}, hereafter 
Z02). The absorber undergoes rapid daily changes of $N_{\rm H}$ 
\citep{Puccetti2007}. Also, a narrow Fe $K_{\alpha}$ line with the equivalent 
width of 50--200 eV is observed (Z02; \citealt{Schurch2003,DeRosa2007}). 

NGC 4151 has frequently been observed by \gro\/ and \sax. The e-folded power law model yields $E_{\rm c}\simeq 50$--200 keV \citep{Johnson1997,DeRosa2007}. Thermal Compton models yield $kT_{\rm e}\simeq 50$--80 keV, $\tau\simeq 1$--2 (\citealt{Johnson1997}; Z02). The strength of Compton reflection is moderate, $R \approx 0.4$ (e.g., \citealt{DeRosa2007}). 

NGC 4151 was observed by \integral\/ 8 times as a primary or secondary target. The results of the first dedicated observation of 2003 May are given in \citet{Beckmann2005}. The object was found in a spectral state similar to those observed with OSSE, but at the highest flux level ever noticed. There was no significant spectral variability despite $\sim 50$ per cent variations of the flux. A thermal Compton model fitted to the summed spectrum yielded $kT_{\rm e}\simeq 90$ keV, $\tau\simeq 1.3$, $R\simeq 0.7$, $N_{\rm H}\simeq 7 \times 10^{22}$ cm$^{-2}$. During almost all later \integral\/ observations the source was found at moderate to very low flux levels. This has allowed us to study the spectral properties of the dim state of this AGN with unprecedented precision and compare them with the bright state of 2003 May.

\section[]{Observations and Data Reduction}
\label{data}

We use all NGC 4151 data collected by \integral\/ as of 2010 January, see Table \ref{obsint}. Good quality data come from the dedicated NGC 4151 observations made in 2003 May, 2007 January, May and December, and 2008 May. The remaining data are taken from observations of the Coma cluster, Mrk 273, NGC 4736, M51 and Mrk 421, when NGC 4151 was almost always seen at an off-axis angle of 9\degr--15\degr. We select the data with the off-axis angle $<15\degr$ for ISGRI and SPI, and $<3\degr$ for the JEM-X and OMC. The data have been reduced using the Offline Scientific Analysis ({\sc osa}) 7.0 provided by the \integral\/ Science Data Centre \citep{Courvoisier2003}, with the pipeline parameters set to the default values. Since the {\sc osa} v.\ 8.0 of 2009 August brought a major improvement for the JEM-X data analysis, we have used it for these data. The ISGRI and SPI spectra and light curves have been extracted with the standard spectral extraction software, including the catalogue sources NGC 4151, NGC 4051, NGC 4138, Mrk 766, NGC 4258, NGC 4395, Coma Cluster, NGC 5033, Mrk 421 and Mrk 268. The JEM-X spectra have been obtained from mosaic images stacked for a given observation. We use the standard response files for all instruments. The OMC magnitude was converted into flux using the calibration of \citet{Johnson1966}, $F({\rm V}\!=\!0)= 3.92\times 10^{-9}$ erg cm$^{-2}$ s$^{-1}$ \AA$^{-1}$.

\begin{table*}
\centering
\caption{The observation log of \integral. The ISGRI exposure is effective, corresponding to fully coded observations, and the JEM-X one is summed over
all pointings with the off-axis angle $<3\degr$. The last column gives the spectral set as defined in Section \ref{assumptions}; B = bright,
D = dim, M = medium.}
\label{obsint}
\begin{tabular}{@{}cccccccc@{}}
\hline
Revolution & Start time, UTC (MJD)  & End time, UTC (MJD)  & Eff. exposure [s] & JEM-X exposure [s] 
& Spectral set \\
\hline
0036 & 2003-01-29 16:43 (52668.697) & 2003-01-31 08:18 (52670.346) &  11296 & -- & -- \\
0071 & 2003-05-14 12:42 (52773.529) & 2003-05-16 11:39 (52775.485) &   9908 & -- & -- \\
0072 & 2003-05-17 13:22 (52776.557) & 2003-05-18 19:56 (52777.831) &   5376 & -- & -- \\
0073 & 2003-05-20 18:44 (52779.781) & 2003-05-22 10:23 (52781.433) &  20441 & -- & -- \\
0074 & 2003-05-23 08:35 (52782.358) & 2003-05-25 11:42 (52784.488) & 117528 &  91633 & B \\
0075 & 2003-05-25 21:03 (52784.877) & 2003-05-28 05:33 (52787.231) & 153481 & 153481 & B \\
0076 & 2003-05-28 20:51 (52787.869) & 2003-05-29 12:13 (52788.509) &  39533 &  39533 & -- \\
0274 & 2005-01-10 16:05 (53380.670) & 2005-01-12 17:36 (53382.733) &   9146 & -- & -- \\
0275 & 2005-01-13 19:36 (53383.817) & 2005-01-15 04:20 (53385.181) &  11745 & -- & -- \\
0310 & 2005-04-29 16:32 (53489.689) & 2005-04-30 03:46 (53490.157) &   1734 & -- & D \\
0311 & 2005-05-01 08:54 (53491.371) & 2005-05-02 20:28 (53492.853) &   3290 & -- & D \\
0312 & 2005-05-03 22:44 (53493.947) & 2005-05-04 09:57 (53494.416) &   1062 & -- & D \\
0317 & 2005-05-19 18:04 (53509.753) & 2005-05-21 01:09 (53511.048) &  10144 & -- & D \\
0318 & 2005-05-22 01:03 (53512.044) & 2005-05-24 06:37 (53514.276) &   7894 & -- & D \\
0324 & 2005-06-09 17:55 (53530.747) & 2005-06-10 23:12 (53531.967) &   9748 & -- & D \\
0448 & 2006-06-14 09:54 (53900.413) & 2006-06-17 00:03 (53903.002) &  29525 & -- & D \\
0449 & 2006-06-17 10:19 (53903.430) & 2006-06-19 23:47 (53905.991) &  28497 & -- & D \\
0450 & 2006-06-20 09:23 (53906.391) & 2006-06-22 23:29 (53908.978) &  26974 & -- & D \\
0451 & 2006-06-23 09:06 (53909.379) & 2006-06-25 23:10 (53911.965) &  26844 & -- & D \\
0521 & 2007-01-18 16:45 (54118.698) & 2007-01-20 23:28 (54120.978) & 122878 & 122878 & D \\
0522 & 2007-01-21 16:35 (54121.691) & 2007-01-24 06:07 (54124.255) & 123751 & 105848 & D \\
0561 & 2007-05-18 09:31 (54238.397) & 2007-05-20 16:11 (54240.674) & 110696 & 110696 & D \\
0562 & 2007-05-22 00:44 (54242.031) & 2007-05-23 21:34 (54243.899) & 100729 &  78177 & D \\
0563 & 2007-05-24 09:06 (54244.379) & 2007-05-25 06:31 (54245.272) &  46903 &  46903 & D \\
0634 & 2007-12-22 17:53 (54456.745) & 2007-12-25 07:19 (54459.305) & 131811 & 131811 & M \\
0636 & 2007-12-28 17:26 (54462.726) & 2007-12-31 01:53 (54465.078) & 119738 & 100028 & M \\
0678 & 2008-05-02 21:17 (54588.887) & 2008-05-04 22:36 (54590.942) &  24805 & -- & M \\
0679 & 2008-05-05 20:47 (54591.866) & 2008-05-07 07:09 (54593.298) &  17266 & -- & M \\ 
0809 & 2009-05-31 11:31 (54982.480) & 2009-05-31 20:58 (54982.874) &  20019 & 2891 & -- \\
0810 & 2009-06-01 05:49 (54983.242) & 2009-06-03 16:10 (54985.674) & 112086 & 46546 & -- \\
0811 & 2009-06-04 05:37 (54986.234) & 2009-06-06 10:24 (54988.433) & 108501 & 40239 & -- \\
\hline
\end{tabular}
\end{table*}

To better constrain the spectra at low energies, we supplemented the \integral\/ data by all available X-ray observations of NGC 4151 taken since 2003 by \xte, \xmm\/ and \suzaku, see Table \ref{obsother}. The \xte\/\/ PCU2 light curves and spectra and HEXTE spectra were extracted with the {\sc heasoft} 6.5.1, using standard selection criteria. The 5 \xmm\/ observations (denoted here X1--X5) were analyzed with {\sc sas} 8.0.1. We used only the EPIC pn data, and excluded periods of high or unstable background. The spectra X1 and X2 have been found to be compatible with each other; we have therefore added them together. For the single \suzaku\/ observation, the data in the $3\times 3$ mode were reduced with {\sc heasoft} 6.6. The spectra of the front-illuminated CCD, XIS0 and XIS3, were added together, whereas the spectrum of the back-illuminated detector XIS1 was used separately. We also extracted the HXD/PIN spectrum using the standard procedure. Although there have been a number of \chandra\/ NGC 4151 observations, all those without gratings that were public at the time we started the analysis were from before 2000 March.

Tables \ref{obsint}--\ref{obsother} identify the data used for our three main 
spectral sets, bright (B), medium (M), and dim (D), whose selection is based on 
ISGRI flux. The remaining \integral\/ data, not marked with a letter in 
Table \ref{obsint}, were not used for spectral analysis. We have excluded them 
because either there were no corresponding high-quality spectra from X-ray 
satellites (e.g., Revs.\ 0809--0811) and/or their flux level was outside the range assumed by us (e.g., beginning of Rev.\ 0074 and Rev.\ 0076 in the case of the B state). The data sets from the other satellites were not always simultaneous with the \integral\/ data. Thus, we assign them to one of the three sets based again on the flux, using the monitoring data from \swift/BAT (see Sections \ref{variability}, \ref{assumptions} for details of the selection).

\begin{table*}
 \centering
\caption{The observation log for the other X-ray satellites. The exposure times are given for the \xte\/ PCU2 (where the number of pointings added together is shown in parentheses), EPIC pn (\xmm) and XIS (\suzaku) detectors. The last column shows the data set as defined in Table \ref{obsint}, and, in the case of \xmm, its consecutive observation number.}
\label{obsother}
\begin{tabular}{@{}lcccccr@{}}
\hline
Obs. ID & Start time, UTC (MJD) & End time, UTC (MJD) & Exposure & Spectral set \\ 	
\hline
\multicolumn{5}{c}{\it RXTE} \\
80416-01-01-(00--09) & 2003-05-24 05:49 (52783.242) & 2003-05-29 04:18 (52788.179) & 15760 (11) & B \\  
92113-08-(06--09)    & 2006-05-13 02:28 (53868.103) & 2006-06-23 15:42 (53909.654) & 4240 (4)   & D \\
92113-08-(22--35)    & 2006-12-23 12:51 (54092.535) & 2007-06-22 08:57 (54273.373) & 14272 (14) & D \\
\hline
\multicolumn{5}{c}{\it XMM-Newton} \\
0143500101 & 2003-05-25 01:38 (52784.068) & 2003-05-25 06:54 (52784.288) &  5821 & X1, B \\
0143500201 & 2003-05-26 20:35 (52785.858) & 2003-05-27 01:51 (52786.077) & 11389 & X2, B \\
0143500301 & 2003-05-27 15:17 (52786.637) & 2003-05-27 20:33 (52786.856) & 12291 & X3, B \\
0402660101 & 2006-05-16 06:22 (53871.265) & 2006-05-16 17:35 (53871.733) & 27980 & X4, D \\
0402660201 & 2006-11-29 17:20 (54068.722) & 2006-11-30 08:00 (54069.333) & 21468 & X5, M \\ 
\hline
\multicolumn{5}{c}{\it Suzaku} \\
701034010 & 2006-12-18 20:05 (54087.837) & 2006-12-21 09:14 (54090.385) & 124980 & D \\
\hline
\end{tabular}
\end{table*}

\section[]{Variability}
\label{variability}

\begin{figure}
\includegraphics[width=\columnwidth]{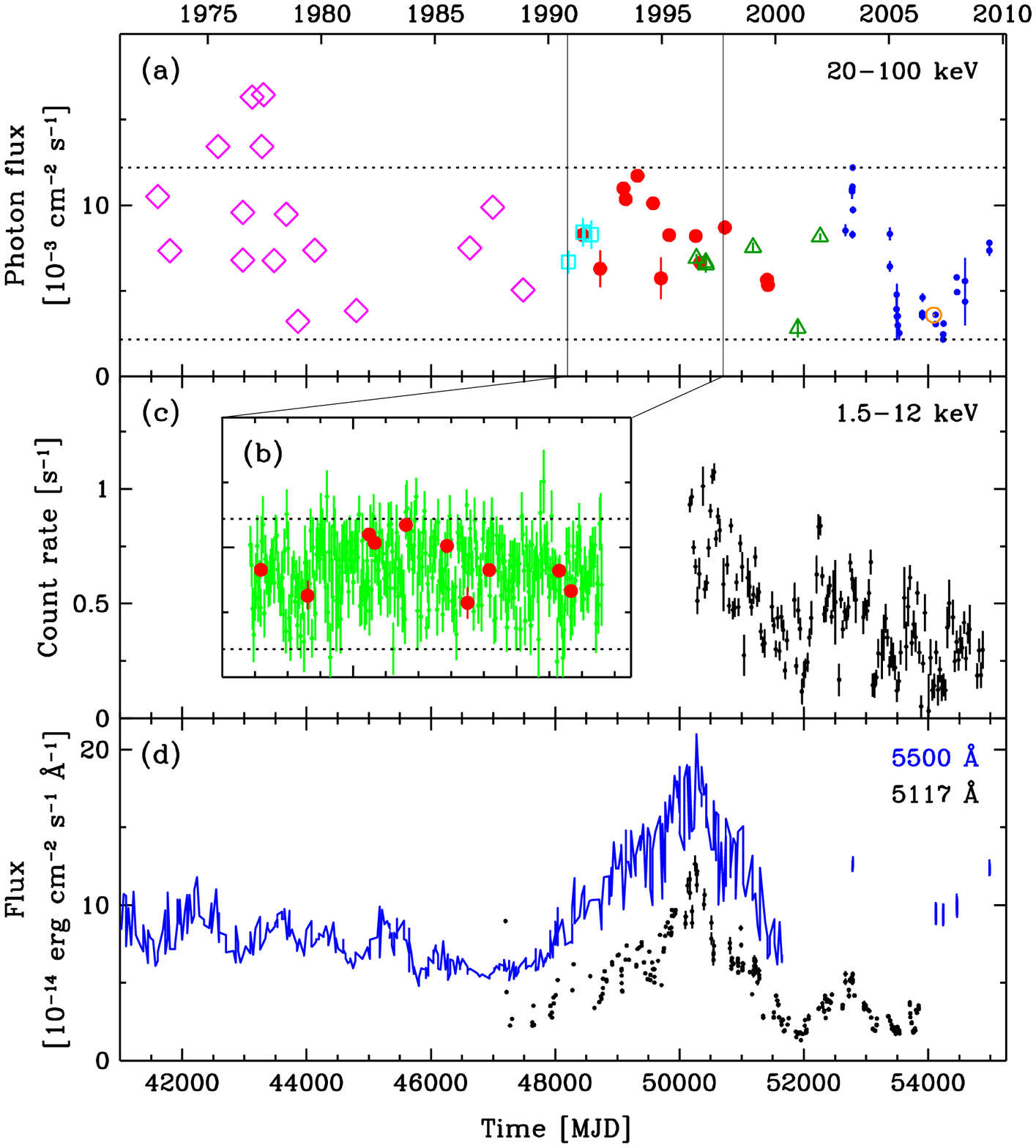}
\caption{Light curves of NGC 4151 since 1972. (a) The 20--100 keV flux from the compilation by \citet{Perotti1991}, magenta diamonds, and from \granat/SIGMA (cyan squares), \gro/OSSE (red filled circles), \sax/PDS (green triangles), \integral/ISGRI (blue dots), and \suzaku/PIN (the orange open circle). The dotted lines show the extrema of the 20--100 keV flux from ISGRI. (b) The 20--100 keV flux from \gro/BATSE (green error bars) together with that from OSSE (red filled circles; same as in a). (c) The \xte/ASM count rate in 30-d bins. (d) The 5500 \AA\ flux (\citealt{Czerny2003}, blue lines), and from \integral/OMC (vertical blue bars at MJD $> 52000$), and the 5117 \AA\ flux (\citealt{Shapovalova2008}, black dots). }
\label{longest}
\end{figure}

NGC 4151 has been observed with X$\gamma$ detectors for almost 40 years. The satellite and balloon results up to 1988 were compiled by \citet{Perotti1991}. Then, it has been observed above 20 keV by \ginga, \granat, \gro, \sax, \integral, \swifte and \suzaku. Fig.\ \ref{longest}(a) shows the 20--100 keV flux observed since 1972 October compared to the optical flux and the \xte/ASM count rate. Because the spectra from earlier observations are not publicly 
available, the 20--100 keV fluxes were determined using the values of the flux
at 35 keV and of the photon index presented in table 2 of \citet{Perotti1991}. 
For this reason, we do not show their uncertainties. Similarly, the \granat\/ fluxes were determined using table 3a of \citet{Finoguenov1995}, where the ART-P and SIGMA spectra are fitted by a power-law model. For all other 
observations, we use the spectra from {\sc heasarc} (\gro/OSSE, \sax/PDS) or spectra extracted by us (\integral/ISGRI, \suzaku/PIN), where the 20--100 keV flux was computed from a power-law model fit (which was done in the 50--150 keV range for the OSSE data) and the uncertainty was determined from the relative error of the summed count rate in the fitted band. Fig.\ \ref{longest}(b) shows the 20--100 keV fluxes from the \gro/BATSE in 7-d bins (together with those from OSSE). They have been obtained from the 20--70 keV fluxes given by \citet{Parsons1998} by multiplying by 1.1 (assuming $\Gamma=1.8$) and then dividing by the normalization factor with respect to OSSE of 1.42 \citep{Parsons1998}.

Fig.\ \ref{longest}(b) gives the 1.5--12 keV count rate of \xte/ASM\footnote{http://xte.mit.edu/asmlc/ASM.html} in 30-d bins. The blue curve in Fig.\ \ref{longest}(c) shows the optical data from the Crimean observatory in the V band (5500 \AA, \citealt{Czerny2003}), which, after MJD 52000, are supplemented by the \integral/OMC data. Since the V band data do not overlap much in time with the ASM data, we also show the 5117 \AA\ fluxes from various observatories given by \citet{Shapovalova2008}. 

The 1.5--12 keV count rate appears well correlated with the optical fluxes, see Fig.\ \ref{longest}. As found by \citet{Czerny2003}, the medium-energy X-ray flux of NGC 4151 varies on time scales $\sim 5$--$10^3$ d, whereas the optical variability is also present on longer time scales. For hard X-rays, a correlation with the optical and the 1.5--12 keV fluxes is less clear. The main peak of the optical emission at MJD $\simeq 49000$--51000 (1993--1998) is reflected in the 1.5--12 keV flux but not in the 20--100 keV one (including the BATSE data). The later data from \sax, \integral\/ and \suzaku\/ show an overall agreement with the optical and softer X-ray fluxes in a sense that the minima at MJD $\sim$52000 and MJD $\sim$54000 and the maximum at MJD $\sim$52800 appear for all these bands. However, the scarce hard X-ray coverage prevents unambiguous conclusions. 

\begin{figure}
\includegraphics[width=\columnwidth]{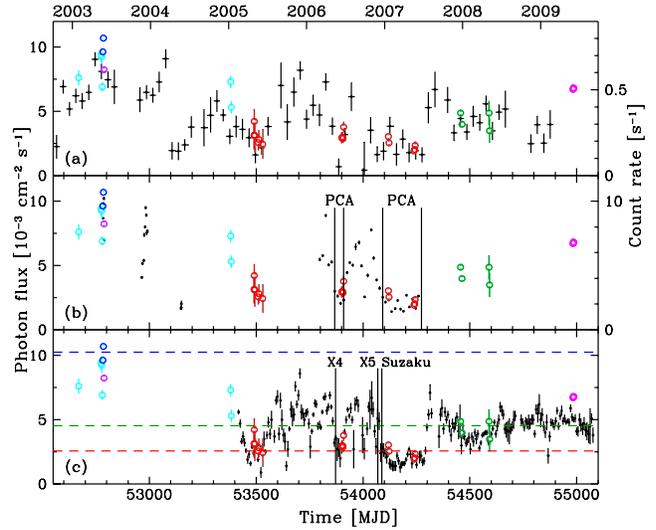}
\caption{Medium and hard X-ray light curves from \integral/ISGRI compared to 
those from \xte\/ and \swift\/ (shown with black circles and error bars). The 
18--50 keV ISGRI flux is shown by open circles on all panels. The colours 
identify the three main states, bright (blue), medium (green), dim (red) and, 
additionally, data excluded from the spectral analysis, Revs.\ 0076 and 
0809--0811 (magenta), and the rest (cyan). (a) The \xte/ASM count rate in 30-d 
bins (right axis). (b) The \xte/PCU2 count rate (right axis). The vertical 
lines delineate the two periods for which the dim state PCA spectrum was 
obtained. (c) Comparison with the \swift/BAT 15--50 keV flux in 7-d bins. The 
vertical lines show the times of the X4 (dim), X5 (medium) \xmm\/ observations 
and that of \suzaku\/ (dim). The horizontal dashed lines show the average flux 
levels of ISGRI bright, medium and dim states. 
}
\label{long}
\end{figure}

The results shown in Fig.\ \ref{longest}(a--b) show that the hard X-ray emission of NGC 4151 has varied in well defined limits over the last 39 years. Almost all fluxes are within the extreme 20--100 keV fluxes from ISGRI (with the maximum in 2003 May and the minimum in 2007 May). There are four flux measurements from older (1975--77) observations above this maximum. However, they are relatively doubtful, corresponding to marginal ($\approx 2$--$3\sigma$) detections at $\ga$ 40 keV. Thus, ISGRI 20--100 keV flux range appears very close to the corresponding overall actual range. This is also supported by the BATSE data, Fig.\ \ref{longest}(b), varying within this range. We also notice that NGC 4151 was rarely seen at low hard X-ray fluxes before the \integral\/ launch.

Fig.\ \ref{long} compares ISGRI 18--50 keV fluxes with the contemporaneous data in the medium and hard X-ray bands from \xte\/ and \swift. The ISGRI fluxes are well correlated with those from both \xte\/ and the \swift/BAT\footnote{http://swift.gsfc.nasa.gov/docs/swift/results/transients}. In particular, we see a good correlation with the 1.5--12 keV rate in spite of the variable absorption affecting a lower part of this band. Fig.\ \ref{long} also shows that a large fraction of the \integral\/ observations happened during periods with very low hard X-ray flux. Fig.\ \ref{long} also identifies the \integral\/ observations used for the spectral sets (Table \ref{obsint}). 

\begin{figure}
\includegraphics[width=\columnwidth]{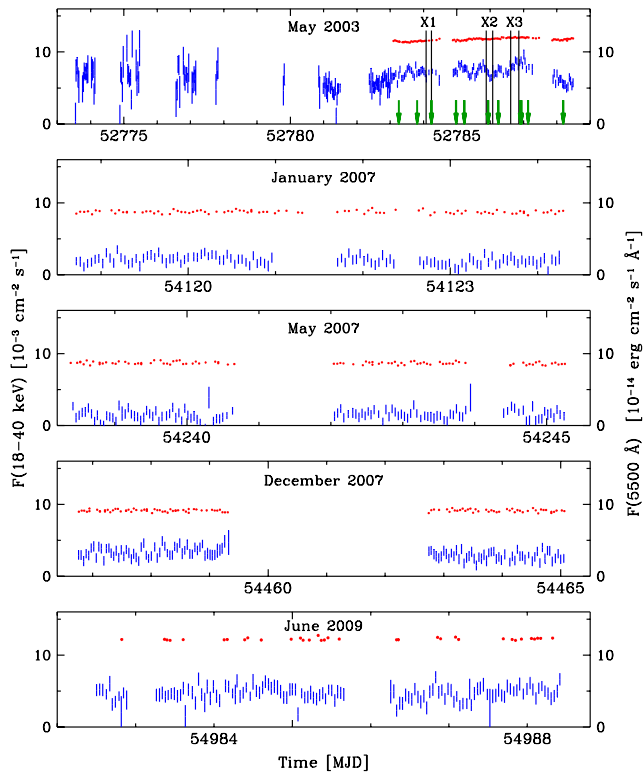}
\caption{Illustration of short time-scale variability. The blue bars (left axis) and the red dots show the 18--40 keV ISGRI flux and the optical OMC flux, respectively, during the dedicated \integral\/ observations. The vertical lines and the green arrows in the top panel show the times of the bright-state X1--X3 \xmm\/ and 11 \xte/PCA observations, respectively. The data used for the bright \integral\/ spectrum are those in the top panel simultaneous with the OMC observations except those around MJD 52788 (Rev.\ 0076), which have markedly lower fluxes. }
\label{short}
\end{figure}

The BAT light curve, Fig. \ref{long}(c), shows that the hard X-ray flux varies by a factor of a few on a week-time scale. When the source is sufficiently 
bright, it is possible for ISGRI to monitor hour time-scale variability. This was the case for the bright state of 2003 May, and the source showed such variability, see Fig.\ \ref{short}. The shown ISGRI and OMC light curves were extracted with a time bin equal to a single pointing duration (10 m--2.5 h, typically $\sim 1$ h). When observed in dimmer states, ISGRI flux remains constant within the measurement errors during a given continuous observation. The optical flux appears constant on hour--day time scales. Still, it varies on longer time scales, and it increases by $\sim$30 per cent when the source is bright in X-rays (2003 May, 2009 June). 

We have extracted the bright (X1--X3) \xmm/EPIC pn light curves in several energy bands in 100-s time bins, and compared with the OMC 100-s light curves. No variability is seen in either light curves on time scales $\la 1$ h. The \xmm\/ light curves vary slowly in good agreement with the trends observed for ISGRI (and also JEM-X) fluxes. 

\begin{figure}
\includegraphics[width=\columnwidth]{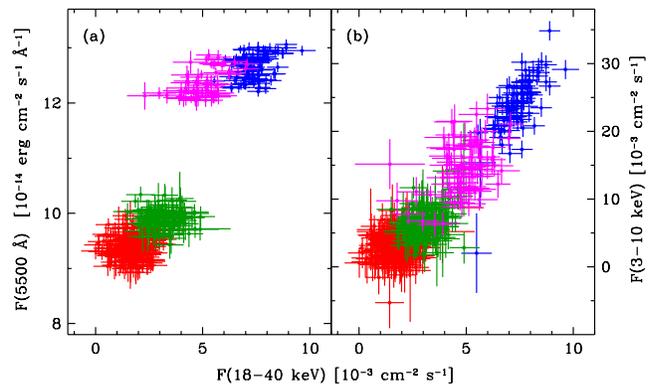}
\caption{(a) The OMC 5500 \AA\ flux vs.\ the ISGRI 18--40 keV flux. (b) The JEM-X 3--10 keV flux vs.\ the ISGRI flux. Each cross corresponds to a single \integral\/ pointing. The colours identify the three main states, bright (blue), medium (green), dim (red), and, additionally, Revs.\ 0076, 0809--0811 (magenta). }
\label{correl}
\end{figure}

The \integral\/ data are suitable for correlation studies because they provide 
simultaneous data in the optical, medium and hard X-ray bands. Here, we use all 
the observations with the off-axis angle $<3\degr$ for which OMC and
JEM-X data are available. They include also part of the observations not 
used in our spectral analysis, namely those identified with magenta circles in Fig.\ \ref{long}. Fig.\ \ref{correl} shows the correlations between the 18--40 keV flux and those fluxes in the 3--10 keV and V bands. Both show strong 
correlations, but the correlation patterns are different. The optical fluxes 
fall into two separate regions, within each they are correlated with the hard 
X-rays. Significant rank-order probabilities ($>\! 0.999$) of the ISGRI/OMC 
correlation are found separately for each of the regions, but also and for the 
entire data set. The observed bimodality may result from strong optical 
long-term variability between the X-ray observations. On the other hand, 
the 3--10 keV and 18--40 keV fluxes exhibit a single, extremely strong, linear 
correlation. Statistical uncertainty blurs somewhat the correlation for low 
fluxes. This result indicates that during all \integral\/ observations the slope 
of the X-ray spectrum of NGC 4151 below 40 keV remains approximately constant.

\section[]{Spectral analysis}
\label{spectral}

\subsection{Assumptions and selection of spectral sets}
\label{assumptions}

Spectral fitting is performed with {\sc xspec} 11.3 \citep{Arnaud1996}. Errors are given for 90 per cent confidence level for a single parameter, $\Delta 
\chi^{2} = 2.7$. The luminosity and distance, $D=13.2$ Mpc, are for $H_{0} = 75$ km s$^{-1}$ Mpc$^{-1}$. We use the elemental abundances of \citet{Anders1982} and the photoelectric absorption cross-sections from \citet{Balucinska1992}. The Galactic column density is set to $N_{\rm H}^{\rm G}= 2.1\times 10^{20}$ cm$^{-2}$. The assumed inclination angle of the reflector and the hot plasma (for anisotropic geometries) is $i=45\degr$ \citep{Das2005}. The reflector is assumed to be neutral, and relativistic broadening of reflected spectra is neglected. Still, in some cases we test for the effects of either a variable Fe abundance of the intrinsic absorber and the reflector, a variable $i$, or an ionization of the reflector.

The main assumption used in our selection of spectral sets is that the state of the emitting hot plasma is determined by the hard X-ray flux. This is justified by our finding that the shape of ISGRI spectra for a given revolution is almost constant at a given hard X-ray flux. As an additional constraint, we do not split \integral\/ data from a single revolution into different spectral sets. Based on this criterion, we select three average ISGRI spectra, as specified in Table \ref{obsint} and Fig.\ \ref{long}. These spectra are accompanied by spectra taken with the other instruments. The state is assigned to each of them on the basis of the light curves presented in Figs.\ \ref{long}--\ref{short}. In particular, the high quality of the \swift/BAT data allows us to determine the hard X-ray flux for periods with no \integral\/ observations, and thus to assign the correct state to the existing lower-energy data sets (Table \ref{obsother}). On the other hand, the data from \xte\/ determine by themselves the hard X-ray state. 

The ISGRI bright (B) and dim (D) spectra are of relatively good quality, and correspond to the extrema of the measured hard X-ray flux. For the flux range in-between, we identify two flux ranges. However, we had no corresponding low-energy data sets for the upper of them. Thus, we use the lower of them, denoted as medium (M). The fitted data sets are as follows.

The bright state, B. We use the data from 2003 May made at low off-axis angles, 
as shown in the top panel of Fig.\ \ref{short}. The spectrum includes a part of 
Rev.\ 0074 (without its beginning) and Rev.\ 0075. For these data, we have 
simultaneous \xte\/ and \xmm\/ observations, see Table \ref{obsother}. 
The fitted spectral set consists of ISGRI spectrum, the X1+X2 and X3 \xmm\/ 
spectra, and the \xte\/ PCU2 spectrum summed over 11 observations.

The medium state, M. We sum ISGRI data from Revs.\ 0634, 0636, 0678--0679, and use the summed JEM-X 1 spectrum from Revs.\ 0634 and 0636, and the X5 \xmm\/ spectrum. 

The dim state, D. The summed ISGRI spectrum is from Revs.\ 0310--0563. We use the \xte\/ PCU2, the X4 \xmm\/ EPIC pn, and the \suzaku\/ XIS0,3 and XIS1 spectra. 

We have also investigated the effect of including other available spectra for each of the data set, whose resulting sets we denote 'extended'. The extended set B also includes the spectra from the \integral/JEM-X 2 and SPI, and the \xte/HEXTE clusters 0, 1. The extended set M also includes the SPI spectrum. The extended data set D also includes two \integral\/ JEM-X 1 summed spectra, from Revs.\ 0521--0522 and 0561--0563, the \suzaku\/ PIN spectrum, the \xte\/ HEXTE cluster 1 spectrum, and the corresponding SPI spectrum. We discuss the effect of using the extended data sets in Section \ref{Compton} below.

In order to limit the complexity of the fitted models, we have decided not to use the X-ray spectra at $E< 2.5$ keV. A two-component partial absorption was then sufficient to model absorption of the continuum. Using the X-ray spectra in the 0.1--2.5 keV band would have required to add a number of spectral lines as well as a soft excess component. As we have tested, including that low-energy band would not affect our determination of the parameters of the hot plasma and of Compton reflection.

Thus, the EPIC pn, XIS0+XIS3, and XIS1 spectra were used in the 2.5--11.3 keV, 
2.5--9.5 keV and 2.5--9.0 keV bands, respectively. The PCA/PCU2 and JEM-X 
spectra were fitted in the 3--16 keV and 4--19 keV bands, respectively. The 
HXD/PIN spectrum was used in the 19--60 keV band because the $<$19-keV data were 
showing a strong excess. For ISGRI, data below 20 keV were always excluded, 
whereas the high-energy limit was in the 180--200 keV range, depending on our 
$3\sigma$ detection threshold used to select good spectral channels. The ISGRI 
spectra in the staring mode from Revs.\ 0074--75 show an excess of $\sim$10 per 
cent below 23 keV, whose origin remains unknown. It could be related to a 
specific setting of the low-energy threshold in the early period of the mission. Thus, we used only the data at $>23$ keV in this case. The HEXTE spectra were used in the 13--160 keV band for the bright state and in 23--110 keV for the dim state. The SPI spectra were fitted in the 23--200 keV band. 

While fitting a given spectral set, the parameters of intrinsic continua are the same for all included spectral data, allowing only the normalization of the model to vary between them. This reflects our assumption that the hard X-ray flux determines the intrinsic state of the source, even for non-simultaneous data. On the other hand, absorption and the Fe K emission may vary quickly \citep{Puccetti2007}, and we cannot expect their parameters to be the same for different low-energy spectra associated with the same spectral set. Thus, we have allowed them to vary between different spectra. Our assumed absorber consists of one fully covering the source with the column density of $N_{\rm H}^{\rm i}$, and the ionization parameter, $\xi \equiv 4\upi F_{\rm i}/n$, where $F_{\rm i}$ is the 5 eV--20 keV irradiating flux and $n$ is the density of the reflector. For the low-resolution detectors, PCA and JEM-X, we assumed $\xi=0$. For the high-resolution detectors, EPIC pn and XIS, we allowed $\xi>0$, and, in addition, we included partial-covering neutral absorption, with the column density of $N_{\rm H}^{p}$ covering a fraction of $f^{\rm p}$ of the flux. For the ionized absorption, we use the model by \citet{Done1992}. Although it is not applicable to highly ionized plasma, it is sufficiently accurate at low/moderate ionization, such as that found in our data. Absorption for ISGRI, HEXTE, SPI and PIN spectra is only marginally important, and we thus used only the neutral fully-covering $N_{\rm H}^{\rm i}$ at the values fitted to the corresponding PCA or JEM-X spectra. The Fe K$\alpha$ emission is modelled by a Gaussian line, with the centre energy, width, photon flux and equivalent width of $E_{\rm Fe}$, $\sigma_{\rm Fe}$, $I_{\rm Fe}$ and $w_{\rm Fe}$, respectively. Fitting the dim state spectra, we have found that adding an Fe K$\beta$ line (at $\approx$7.1 keV) was significantly improving the fit for the EPIC pn and XIS spectra.  

\subsection{The cut-off power-law model}
\label{cutoff}

We first fit an e-folded power-law model including a Compton reflection component. This allows us to compare our results to other published ones. We use the {\sc pexrav} model \citep{Magdziarz1995}, and our other assumptions are specified above. The results are shown in Table \ref{pl_fits}, where we do not show the fitted absorber and line parameters since they are similar to those for our main model of thermal Comptonization, see Section \ref{Compton} below. 

Our results regarding $\Gamma$ and $R$ are similar to those of earlier works. However, a significant difference is found for the e-folding energy, $E_{\rm c}$, which we find to be high, and compatible with no cut-off for two of our states, see Table \ref{pl_fits}. A finite $E_{\rm c}$ is found only for the bright state. Even then, we obtain values well above those reported before, which are 30--45 keV (P01), 80--180 keV \citep{DeRosa2007} and $\sim$210 keV (Z02). As discussed in \citet{Zdziarski2003}, an exponential cut-off is much shallower than that characteristic of thermal Comptonization, and not sharp enough to model the spectral high-energy cut-offs observed in Seyferts, in particular in NGC 4151. Thus, our high values of $E_{\rm c}$ result from the good quality of the data below the cut-off, which dominate the statistics and force an approximately straight power law to continue to just below the beginning of the cut-off.

\begin{table}
\setlength{\tabcolsep}{1.2mm}
\begin{center}{
\caption{Spectral fitting results for the e-folded power-law model including reflection. The {\sc xspec} model is {\sc constant*wabs*absori*zpcfabs(pexrav\ +zgauss)}, and the normalization, $K$, is given by the photon flux at 1 keV in units of $10^{-2}$ keV$^{-1}$ cm$^{-2}$ s$^{-1}$. }
\small
\label{pl_fits}
\begin{tabular}{cccccc}
\hline
State & $E_{\rm c}$ [keV] & $\Gamma$ & $R$ & $K$ & $\chi ^{2}$/d.o.f. \\
\hline
Bright & 264$_{-26}^{+48}$ & 1.71$_{-0.01}^{+0.06}$ & 0.45$_{-0.05}^{+0.12}$ & 8.7$_{-0.3}^{+1.1}$ & 3179/3414 \\
Medium & $>$ 1025 & 1.81$_{-0.03}^{+0.05}$ & 0.6$_{-0.3}^{+0.3}$ & 4.6$_{-0.7}^{+1.4}$ & 1554/1599 \\
Dim & $>$ 1325 & 1.81$_{-0.01}^{+0.01}$ & 0.92$_{-0.05}^{+0.05}$ & 2.24$_{-0.02}^{+0.01}$ & 3338/3384 \\
\hline
\end{tabular}
}\end{center}
\end{table}

\subsection{Comptonization model}
\label{Compton}

\begin{table*}
\begin{minipage}{172mm}
\begin{center}
\caption{{\bf (a)} Spectral fitting results with thermal Comptonization as the main continuum. The {\sc xspec} model is {\sc constant*wabs*absori*zpcfabs*(compps\ +zgauss)}. See Section \ref{Compton} for the definitions of the parameters. $L_{\rm X\gamma}$ and $L_{\rm X\gamma}'$ are the bolometric model luminosity (in units of $10^{43}$ erg s$^{-1}$) with and without the reflection component, respectively, and only the values for the geometry 0 are given as the values for other geometries are similar. }
\small
\label{comp_results}
\begin{tabular}{lccccccccc}
\hline
State & $kT_{\rm e}$ [keV] & $y$ & $\tau$ & $R$ & $K\,[10^8]$ & $L_{\rm X\gamma}$ & $L_{\rm X\gamma}'$ &Geometry & $\chi ^{2}$/d.o.f. \\
\hline
Bright & 54$_{-5}^{+7}$ & 1.10$_{-0.02}^{+0.02}$ & 2.6$_{-0.3}^{+0.3}$ & 0.40$_{-0.06}^{+0.07}$ & 5.2$_{-0.6}^{+1.4}$ & 5.18 & 4.61 & sphere (0) & 3165/3414 \\
& 62$_{-7}^{+7}$ & 0.60$_{-0.01}^{+0.01}$ & 1.3$_{-0.1}^{+0.1}$ & 0.55$_{-0.07}^{+0.08}$ & 7.3$_{-0.9}^{+1.9}$ & & & slab (1) &  3166/3414 \\
& 62$_{-7}^{+9}$ & 0.61$_{-0.01}^{+0.01}$ & 1.3$_{-0.2}^{+0.1}$ & 0.55$_{-0.06}^{+0.09}$ & 3.9$_{-0.7}^{+0.4}$ & & & slab ($-1$) & 3166/3414 \\
& 73$_{-9}^{+14}$ & 1.06$_{-0.02}^{+0.02}$ & 1.9$_{-0.2}^{+0.4}$ & 0.47$_{-0.07}^{+0.05}$ & 12.0$_{-1.6}^{+2.9}$ & & & cylinder (2) & 3168/3414 \\
& 73$_{-13}^{+13}$ & 1.06$_{-0.01}^{+0.03}$ & 1.9$_{-0.3}^{+0.3}$ & 0.51$_{-0.9}^{+0.5}$ & 5.4$_{-1.1}^{+1.4}$ & & & cylinder ($-2$) & 3168/3414 \\
& 73$_{-7}^{+16}$ & 1.21$_{-0.02}^{+0.02}$ & 2.1$_{-0.6}^{+0.4}$ & 0.48$_{-0.06}^{+0.06}$ & 11.9$_{-1.4}^{+3.2}$ & & & hemisphere (3) & 3168/3414 \\
& 73$_{-9}^{+15}$ & 1.21$_{-0.02}^{+0.03}$ & 2.1$_{-0.4}^{+0.3}$ & 0.50$_{-0.06}^{+0.08}$ & 4.7$_{-1.0}^{+1.5}$ & & & hemisphere ($-3$) & 3168/3414 \\
& 61$_{-5}^{+7}$ & 0.80$_{-0.02}^{+0.01}$ & 1.7$_{-0.2}^{+0.1}$ & 0.38$_{-0.05}^{+0.04}$ & 1.2$_{-0.1}^{+0.1}$ & & & sphere (4) & 3172/3414 \\
& 57$_{-7}^{+13}$ & 0.86$_{-0.01}^{+0.01}$ & 1.9$_{-0.4}^{+0.2}$ & 0.39$_{-0.05}^{+0.06}$ & 6.4$_{-0.6}^{+0.6}$ & & & sphere ($-4$) & 3169/3414 \\
& 57$_{-5}^{+10}$ & 0.86$_{-0.01}^{+0.01}$ & 1.9$_{-0.3}^{+0.2}$ & 0.38$_{-0.05}^{+0.06}$ & 4.9$_{-0.4}^{+1.5}$ & & & sphere ($-5$) & 3166/3414 \\
\hline
Medium & 128$_{-24}^{+74}$ & 1.02$_{-0.13}^{+0.08}$ & $1.0_{-0.3}^{+0.5}$ & $< 0.41$ & 3.9$_{-1.0}^{+0.4}$ &2.26 & 2.16 & sphere (0) & 1554/1597 \\
\hline
Dim & 190$_{-12}^{+13}$ & 0.98$_{-0.01}^{+0.01}$ & 0.66$_{-0.05}^{+0.04}$ & 0.75$_{-0.04}^{+0.04}$ & 1.54$_{-0.01}^{+0.01}$ &1.47 & 1.19 & sphere (0) & 3337/3384 \\
& 200$_{-8}^{+7}$ & 0.49$_{-0.01}^{+0.01}$ & 0.31$_{-0.01}^{+0.01}$ & 1.01$_{-0.05}^{+0.05}$ & 2.04$_{-0.02}^{+0.01}$ & & & slab (1) & 3335/3384 \\
& 188$_{-7}^{+6}$ & 0.48$_{-0.01}^{+0.01}$ & 0.33$_{-0.01}^{+0.01}$ & 1.04$_{-0.05}^{+0.05}$ & 1.20$_{-0.01}^{+0.01}$ & & & slab ($-1$) & 3350/3384 \\
& 210$_{-14}^{+12}$ & 1.05$_{-0.01}^{+0.01}$ & 0.64$_{-0.04}^{+0.04}$ & 0.79$_{-0.04}^{+0.04}$ & 2.85$_{-0.01}^{+0.02}$ & & & cylinder (2) & 3334/3384 \\
& 215$_{-13}^{+12}$ & 1.01$_{-0.01}^{+0.01}$ & 0.60$_{-0.03}^{+0.04}$ & 0.86$_{-0.05}^{+0.05}$ & 1.73$_{-0.01}^{+0.01}$ & & & cylinder ($-2$) & 3336/3384 \\
& 227$_{-14}^{+10}$ & 1.21$_{-0.02}^{+0.01}$ & 0.68$_{-0.03}^{+0.04}$ & 0.84$_{-0.04}^{+0.04}$ & 2.89$_{-0.02}^{+0.01}$ & & & hemisphere (3) & 3334/3384 \\
& 220$_{-11}^{+13}$ & 1.14$_{-0.01}^{+0.01}$ & 0.66$_{-0.04}^{+0.03}$ & 0.90$_{-0.05}^{+0.05}$ & 1.71$_{-0.01}^{+0.02}$ & & & hemisphere ($-3$) & 3336/3384 \\
& 186$_{-9}^{+11}$ & 0.66$_{-0.01}^{+0.01}$ & 0.45$_{-0.03}^{+0.02}$ & 0.77$_{-0.04}^{+0.04}$ & 0.866$_{-0.004}^{+0.004}$ & & & sphere (4) & 3341/3384 \\
& 191$_{-17}^{+13}$ & 0.86$_{-0.01}^{+0.01}$ & 0.58$_{-0.04}^{+0.05}$ & 0.77$_{-0.04}^{+0.04}$ & 1.56$_{-0.01}^{+0.01}$ & & & sphere ($-4$) & 3337/3384 \\
& 196$_{-13}^{+13}$ & 0.81$_{-0.01}^{+0.01}$ & 0.53$_{-0.04}^{+0.04}$ & 0.77$_{-0.04}^{+0.04}$ & 1.49$_{-0.01}^{+0.01}$ & & & sphere ($-5$) & 3337/3384 \\
\hline
\end{tabular}
\end{center}
\end{minipage}
\end{table*}

\setcounter{table}{3}
\begin{table*}
\begin{minipage}{172mm}
\begin{center}
\caption{{\bf (b)} Continuation of Table \ref{comp_results}(a). Fit results with the thermal Comptonization model (for a sphere, the geometry parameter $=0$) regarding the absorber, line component, and the relative normalization, $C_{\rm ISGRI}$, of a given X-ray spectrum with respect to ISGRI. See Section \ref{assumptions} for the definitions of the other parameters. The units of $N_{\rm H}$, $\xi$, and $I_{\rm Fe}$ are $10^{22}$ cm$^{-2}$, $10^{-2}$ erg cm s$^{-1}$, $10^{-4}$ cm$^{-2}$ s$^{-1}$, respectively, and `f' denotes a fixed parameter. }
\begin{tabular}{llccccccccc}
\hline
State & X-ray spectrum & $C_{\rm ISGRI}$ & $N_{\rm H}^{\rm i}$ & $\xi$ & $N_{\rm H}^{\rm p}$ & $f^{\rm p}$ & $E_{\rm Fe}$ [keV] & $\sigma_{\rm Fe}$ [keV] & $I_{\rm Fe}$ & $w_{\rm Fe}$ [eV] \\
\hline
Bright & EPIC pn (X1+X2) & 0.92$_{-0.03}^{+0.03}$ & 4.9$_{-0.7}^{+0.9}$ & 10$_{-6}^{+79}$ & 15.8$_{-3.3}^{+3.9}$ & 0.39$_{-0.07}^{+0.09}$ 
& 6.400$_{-0.011}^{+0.011}$ & 0.07$_{-0.01}^{+0.02}$ & 3.0$_{-0.3}^{+0.3}$ & { 83$_{{{-9}}}^{{{+7}}}$ } \\
& EPIC pn (X3)    & 1.08$_{-0.03}^{+0.03}$ & 3.0$_{-1.8}^{+1.5}$ & 64$_{-61}^{+47}$ & 9.5$_{-1.1}^{+4.0}$ & 0.55$_{-0.22}^{+0.18}$ 
& 6.393$_{-0.012}^{+0.013}$ & 0.07$_{-0.01}^{+0.01}$ & 2.8$_{-0.3}^{+0.3}$ & { 76$_{{{-9}}}^{{{+9}}}$ } \\
& PCA PCU2 (B)    & 1.17$_{-0.03}^{+0.03}$ & 7.4$_{-0.4}^{+0.4}$ & 0f & 0f & -- 
& 6.06$_{-0.09}^{+0.08}$ & 0.36$_{-0.13}^{+0.15}$ & 6.8$_{-0.8}^{+1.3}$ & { 169$_{{{-20}}}^{{{+32}}}$ } \\
\hline
Medium & EPIC pn (X5) & 0.83$_{-0.06}^{+0.08}$ & 7.6$_{-0.4}^{+0.5}$ & 4$_{-3}^{+11}$ & 22.6$_{-1.5}^{+4.4}$ & 0.50$_{-0.04}^{+0.05}$ 
& 6.388$_{-0.007}^{+0.007}$ & 0.06$_{-0.01}^{+0.01}$ &  2.7$_{-0.3}^{+0.5}$ & { 149$_{{{-12}}}^{{{+27}}}$ } \\
& JEM-X (M)    & 0.74$_{-0.06}^{+0.06}$ & 13.0$_{-3.3}^{+3.7}$ & 0f & 0f & -- 
& -- & -- & -- & -- \\
\hline
Dim & EPIC pn (X4) & 1.17$_{-0.01}^{+0.01}$ & 6.36$_{-0.15}^{+0.14}$ & 4$_{-3}^{+30}$ & 23.5$_{-0.7}^{+0.6}$ & 0.69$_{-0.01}^{+0.01}$ 
& 6.398$_{-0.005}^{+0.004}$ & 0.06$_{-0.01}^{+0.01}$ & 2.3$_{-0.1}^{+0.1}$ & { 271$_{{{-13}}}^{{{+13}}}$ }  \\
& XIS0,3 (D)   & 1.11$_{-0.01}^{+0.01}$ & 8.03$_{-0.16}^{+0.12}$ & 0.7$_{-0.4}^{+0.9}$ & 28.8$_{-1.0}^{+0.9}$ & 0.50$_{-0.01}^{+0.01}$ 
& 6.384$_{-0.003}^{+0.003}$ & 0.03$_{-0.01}^{+0.01}$ & 2.2$_{-0.1}^{+0.1}$ & { 291$_{{{-13}}}^{{{+13}}}$ } \\
& XIS1 (D)    & 1.13$_{-0.01}^{+0.01}$ & \multicolumn{4}{c}{fitted together with XIS0,3} & 6.412$_{-0.004}^{+0.005}$ 
& 0.04$_{-0.01}^{+0.01}$ & 2.3$_{-0.1}^{+0.1}$ & { 280$_{{{-13}}}^{{{+13}}}$ } \\
& PCA PCU2 (D) & 1.16$_{-0.01}^{+0.01}$ &  13.6$_{-0.5}^{+0.5}$ & 0f & 0f & -- 
& 6.4f & 0f & 2.1$_{-0.3}^{+0.3}$ & { 303$_{{{-39}}}^{{{+39}}}$ } \\
\hline
\end{tabular}
\end{center}
\end{minipage}
\end{table*}

To model thermal Comptonization, we use the {\sc compps} model \citep{Poutanen1996} in {\sc xspec}. It models Compton scattering in a plasma cloud of given temperature, $T_{\rm e}$, and Thomson optical depth, $\tau$, for a number of geometries and locations of the seed photon sources. Since $T_{\rm e}$ and $\tau$ are strongly intrinsically anticorrelated, we also use the Compton parameter (e.g., \citealt{Rybicki1979}), $y\equiv 4(kT_{\rm e}/m_{\rm e} c^2)\tau$ (where $m_{\rm e}$ is the electron mass) as the second fitting parameter instead of $\tau$. The advantage of this choice is that $T_{\rm e}$ and $y$ are almost orthogonal; $y$ determines closely the power-law slope at low energies whereas $kT_{\rm e}$ determines the position of the high-energy cut-off. The seed photons are assumed to have a disc blackbody distribution ({\sc diskbb} in {\sc xspec}, \citealt{Mitsuda84}) with the maximum temperature of $kT_{\rm bb}=10$ eV. The model normalization, $K$, is that of the seed disc blackbody model as defined in {\sc xspec}, see Section \ref{xray}.

Among the geometries included in {\sc compps}, we consider here four spherical cases; an approximate treatment of radiative transfer using escape probability (which is denoted in {\sc compps} by the geometry parameter $=0$), a sphere with central soft photons (geometry 4), homogeneously distributed seed photons (geometry $-4$) and seed photons distributed according to the diffusion-equation eigen-function, $\propto \sin(\upi\tau'/\tau)/ (\upi\tau'/\tau)$, where $0\leq \tau'\leq\tau$ ($-5$). Then, we consider a slab with the seed photons either at its bottom (geometry 1) or distributed homogeneously (geometry $-1$). Also, we consider the hot plasma in the shape of either a cylinder with the height equal to its radius, or a hemisphere, with the seed photons being either at its bottom (2, 3, respectively) or homogeneously distributed ($-2$, $-3$, respectively). 

The spectral data of NGC 4151 studied in this paper are of high quality and represent the largest set ever collected for this source. This gives us a possibility to test how precise information about X$\gamma$ emission from brightest Seyferts can be achieved with the current satellites. Therefore, we extensively test various variants of the Comptonization model for our two best-quality states, bright and dim. This could, in principle, give us some indications regarding the actual geometry of the source. Our results are presented in Tables \ref{comp_results}(a) and (b), regarding the parameters of the Comptonizing plasma and the strength of reflection (which are the same for all used detectors within a given state), and of the absorber and the Fe line (which are specific to a given X-ray detector within each of the states), respectively. The results in Table \ref{comp_results}(b) and the entries for the bolometric model luminosity, $L_{\rm X\gamma}$, in Table \ref{comp_results}(a)
are given only for the case of spherical geometry calculated using escape probability formalism (geometry 0). 

We find the thermal Comptonization model to provide very good fits to the data.  
The values of the reduced $\chi^2$ are $\la 1$, which, however, appears not to 
be due to an overly complex model. In particular, the high-energy continua are 
determined only by three parameters, $y$, $R$, $kT_{\rm e}$. For the bright 
state, Comptonization provides a much better fit than the e-folded power law, 
with $\Delta \chi^2$ of up to $-14$. However, any trends seen for the geometry 
of the Comptonizing plasma are rather weak. In the case of the bright state, 
equally good models can be obtained with either a slab or a sphere. Also, models 
with seed photons being distributed within the source are of similar quality as 
those with localized seed photons. On the other hand, the differences between 
the models are slightly stronger in the case of the dim state. In 
particular, the slab model with seed photons at its bottom is somewhat 
better than that with the seed photons distributed throughout it. This can be a 
hint that the actual source has the seed photons external to the source rather 
than internal (which would be the case for the dominant synchrotron seed 
photons). We also note that the fitted values of $kT_{\rm e}$ are rather 
insensitive to the assumed geometry in both the bright and dim states. Our 
results for the bright state are similar to those of previous Comptonization 
fits, in particular to those of Z02, who obtained $kT_{\rm e}= 73^{+34}_{-29}$ 
keV, $y=0.88^{+0.12}_{-0.11}$, and $R=0.60^{+0.24}_{-0.21}$ at 
$L_{\rm X\gamma}\simeq 3.9\times 10^{43}$ erg s$^{-1}$ using a broad-band 
X-ray spectrum from \asca\/ and OSSE. 

\begin{figure}
\includegraphics[width=\columnwidth]{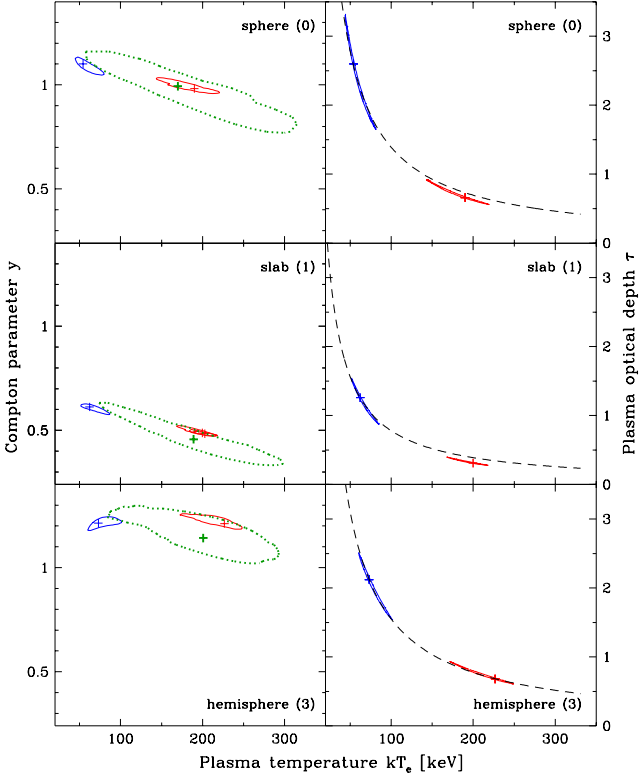}
\caption{The 90 per cent confidence regions for the Compton $y$ parameter (left) and the optical depth, $\tau$ (right), vs.\ the temperature, $kT_{\rm e}$ 
for three geometries of the Comptonizing plasma. The solid blue, solid red and green dashed (only for $y$) contours correspond to the bright, dim and medium state, respectively. The dashed curves for $\tau$ vs.\ $kT_{\rm e}$ correspond to $\tau\propto kT_{\rm e}^{-1}$, normalized to the bright-state results.}
\label{contours}
\end{figure}

In all geometries, we see in Table \ref{comp_results}(a) that the Compton $y$ parameter remains approximately constant between the bright and dim states, with only a slight decrease in the dim state. For the medium state, Table \ref{comp_results}(a) gives the results only for the model with escape probabilities, for which the best fit $y$ is almost the same as that in the dim state. These results are illustrated by the confidence contours shown in Fig.\ \ref{contours}. 

Apart from the results given in Table \ref{comp_results}, a reduction of 
$\chi^{2}$ appears when the Fe abundance (assumed equal for the absorber and 
the reflector) is a free parameter. We found a moderate overabundance, 
${1.5\pm 0.2}$ times that of \citet{Anders1982} for the bright state, with 
$\Delta\chi^{2} = -6$, and ${1.3\pm 0.1}$ for the dim state, $\Delta \chi^{2} = -2$. For the \xmm\/ observation in 2000 December, an Fe overabundance by 2--3 was claimed by \citet{Schurch2003} but that fit was done with the \xmm\/ spectrum only and the reflection strength (affecting that determination) was not well constrained.

We also find that both the bright and dim state spectra prefer an inclination angle $>45\degr$, with $\Delta \chi^{2} = -3$ and $-6$, respectively, at $75\degr$. Since this effect is observed for symmetric (sphere) and asymmetric (hemisphere) geometries, we conclude that its main cause is the changing shape of the reflection component. On the other hand, allowing the reflector to be ionized does not improve the fit, and $\xi < 0.2$ erg cm s$^{-1}$ (at an assumed reflector temperature of $10^5$ K). 

The lack of the far-UV and very soft X-ray spectra does not allow us to fit the maximum temperature of the seed disc blackbody photons, $kT_{\rm bb}$. We have found that changing it to either 5 or 20 eV yields no improvement of the fit. This is indeed expected given that the photon energies emitted by the disc are much below the fitted energy range, at which the shape of the Comptonization spectrum has already achieved a power-law shape independent of $kT_{\rm bb}$.

Table \ref{comp_results}(b) also shows the normalization of a given X-ray spectrum relative to that from ISGRI. For the bright state, when the \integral, \xmm\/ and \xte\/ observations were almost contemporary, $C_{\rm ISGRI}$ is very close to 1 and, for both of the \xmm\/ spectra, it follows the flux changes seen in the top panel of Fig.\ \ref{short}. This shows that ISGRI, EPIC pn and PCA detectors are well cross-calibrated. In the case of the two other spectral sets, the observations are not simultaneous. Nevertheless, $C_{\rm ISGRI}$ varies only within the range of 0.74--1.17, confirming the validity of our selection of the spectra for a given flux state. 

\begin{figure}
\includegraphics[width=\columnwidth]{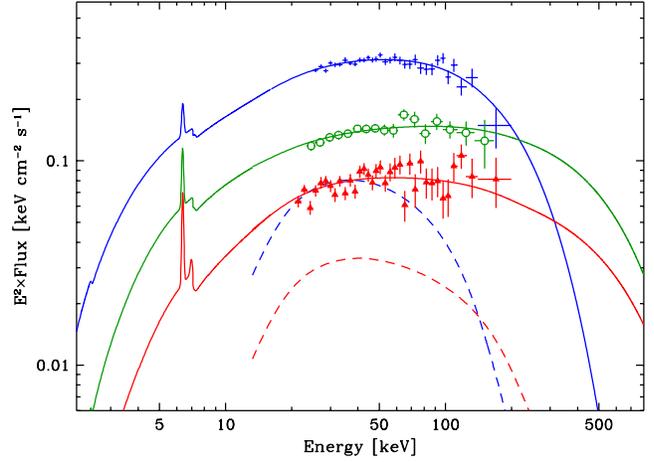}
\caption{The Comptonization model spectra (solid curves; the geometry parameter $=0$) of the three states of the NGC 4151, shown together with ISGRI spectra. The bright, medium and dim spectra are shown by the blue crosses, green circles and red triangles, respectively. The dashed curves show the reflection model components for the bright (upper) and dim state. The fitted JEM-X, \xte, \xmm\/ and \suzaku\/ spectra are not shown for clarity.}
\label{models}
\end{figure}

\begin{figure}
\includegraphics[width=\columnwidth]{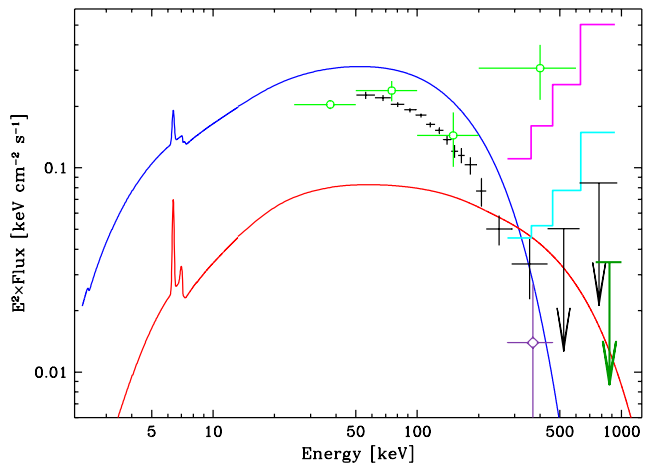}
\caption{The Comptonization models for the extreme states (solid curves, same as in Fig.\ \ref{models}) compared to the average OSSE spectrum (crosses and two $2\sigma$ upper limits). The rightmost upper limit is the COMPTEL upper limit in the 0.75--1 MeV band \citep{Maisack1995}. The histograms give the \integral/PICsIT $3\sigma$ upper limits for 1 Ms (upper) and 10 Ms (lower) effective exposure time. The PICsIT 1.34 Ms average 277--461 keV flux is shown by error bars with a diamond. The SPI spectrum obtained by \citet{Bouchet2008} is shown by error bars with circles. Note that the 200--600 keV data point may be spurious. }
\label{high}
\end{figure}

We have also tested our results using the extended data sets (see Section \ref{assumptions}). However, the quality of the fits and the plasma parameter uncertainty are not improved when they are used. This confirms that our selection of the primary sets was valid. Thus, we do not present here those results.
 
We note that even for such a bright AGN as NGC 4151, \textbf{we have no data at 
energies $\ga 200$ keV, as shown in Fig.\ \ref{models}. Still, for the dim 
state, we have found $kT_{\rm e}\simeq 200$ keV with a statistical uncertainty 
of $\sim 7$ per cent or so. The physical reason for this low uncertainty is the 
observed lack of spectral curvature. In the process of thermal Comptonization, 
a power law slope at $E\la kT_{\rm e}$ appears due to the merging of many 
individual up-scattering profiles, each being curved. The cut-off at $E\ga 
kT_{\rm e}$ is due to the electron energies being comparable to the photon 
ones, in which case a photon no longer increases its energy in a scattering. 
Since we see no cut-off up to $\sim$200 keV, models with a lower $kT_{\rm e}$ 
are ruled out. On the other hand, models with a higher $kT_{\rm e}$ correspond 
to an increase of the photon energy in a single scattering by a factor so large 
that the individual, curved, scattering profiles become clearly visible and the 
spectrum at $E\la kT_{\rm e}$ is no longer a power law. Since the observed 
spectral shape {\it is\/} a power law, this case is also excluded. We note, 
however, that our assumed model corresponds to a single scattering zone. A power 
law spectrum at $E\la 200$ keV may also be produced by a superposition of 
emission regions with different temperatures (smoothing out the resulting 
spectrum), as, e.g., in a hot accretion flow. In this case, the maximum 
temperature of the flow could be higher than 200 keV. In any case, we do see a 
very clear difference in the characteristic Comptonization temperature between 
the bright and dim state.}

We comment here on the very high value of $kT_{\rm e}=315\pm 15$ keV obtained by P01 for the 1999 January \sax\/ spectrum of NGC 4151. The fit of that model appears rather poor when compared with the curvature of the \sax/PDS data above 50 keV, see fig.\ 2(b) in P01. We also find that the PDS spectrum is very similar to the average OSSE spectrum \citep{Johnson1997}, shown in Fig.\ \ref{high} (with only the normalization of the PDS spectrum being about 5 per cent lower). This indicates that the source was in a moderately bright state in 1999 January, for which state the temperature of 315 keV appears highly unlikely. We cannot explain this discrepancy; it may be that $kT_{\rm e}=315\pm 15$ keV represented a spurious local minimum. 

As we see in Fig.\ \ref{high}, our bright-state model spectrum agrees well in shape with the average OSSE spectrum \citep{Johnson1997}. The latter also agrees with the first three points of the average NGC 4151 spectrum from \integral/SPI \citep{Bouchet2008}. However, the SPI 200--600 keV flux is implausibly high, which may reflect a problem with the background estimate. Indeed, the SPI analysis of \citet{Petry2009} yields only upper limits at $E>150$ keV.

We see in Fig.\ \ref{high} that the Comptonization models of the bright and dim 
states cross at $\simeq 300$ keV. Unfortunately, it is impossible to verify it 
with our data. In particular, we cannot test if there is any non-thermal 
component at high energy. Even the average spectrum from OSSE, the most 
sensitive detector up to date at $E\ga 150$ keV, gives only a weak detection 
of NGC 4151 at $E\ga 250$ keV with an $>4$ Ms exposure time. From the shown 
upper limits (including systematic errors) of the \integral/PICsIT, the current 
detector most sensitive at $E\ga 250$ keV \citep{Lubinski2009}, we see it would 
need $>10$ Ms exposure time to detect NGC 4151 in the 300--400 keV range. 
Merging all the suitable \integral\/ data (with the effective exposure time of 
1.34 Ms taken without the bright state, for which the observation was in the 
staring mode, not suitable for the PICsIT), we extracted the PICsIT flux in the 
277--461 keV band. Although it is not a detection (and the shown error bar does 
not include systematic errors), it provides a hint that the average NGC 4151 
emission in this energy range is weak.  

\section[]{Discussion}

\subsection{Reflection and Fe line}
\label{reflector}

\subsubsection{Comparison with previous work}
\label{comparison}

A reflection component in the X-ray spectra of NGC 4151 was first found by
\citet{Zdziarski1996a}, using contemporaneous data from \gro\/ and \ginga,
yielding $R\simeq 0.43$ (using a Comptonization model and assuming $i = 
65\degr$). Then, Z02 found $R\simeq 0.6$ (at $i=17\degr$) using \gro\/ and 
\asca\/ data, and P01 found $R\simeq 0.2$--1.8 using \sax\/ data and a 
reflection model averaged over the viewing angle. A previous analysis 
of \integral\/ bright-state data from 2003 May yielded a larger reflection, 
$R = 0.72\pm 0.14$ \citep{Beckmann2005}, than that determined by us (their 
value of $kT_{\rm e} = 94_{-10}^{+4}$ keV is also different from that found by 
us for the same data). By fitting the ISGRI and JEM-X bright-state spectra
alone, we find results very similar to those with the \xmme and \xtee spectra 
included (Table \ref{comp_results}). Therefore the differences appear to be a 
consequence of the substantial change of the \integral\/ calibration since 
2005. Furthermore, \citet{Beckmann2005} used the SPI spectra obtained in the 
staring observation mode, which can be affected by a large background 
uncertainty. Then, \citet{DeRosa2007} found $R\simeq 0.4$ for several \sax\/ 
spectra and $R\simeq 0$ for the remaining ones, but using an e-folded power-law 
model rather than Comptonization. Then, \citet{Schurch2002} found $R\simeq 1.9$ 
using \xmm\/ data assuming $i=65\degr$, but this result was obtained using the 
2.5--12 keV band only and assuming a fixed $\Gamma = 1.65$.

Regarding Fe K$\alpha$ line, \citet{Zdziarski1996a} found a narrow line with $I_{\rm Fe}\simeq (0.6$--$1.0)\times 10^{-4}$ cm$^{-2}$ s$^{-1}$ using \ginga\/ data, and Z02 found a narrow line at $I_{\rm Fe}\simeq 2.5\times 10^{-4}$ cm$^{-2}$ s$^{-1}$ accompanied by a weaker broad relativistic line using \asca\/ data. The broad component has not been observed later, whereas the narrow line flux measured by \sax\/ \citep{DeRosa2007}, \chandra\/ \citep{Ogle2000} and \xmm\/ \citep{Schurch2003} was found to be in the range (1.3--$4.2)\times 10^{-4}$ cm$^{-2}$ s$^{-1}$. As seen in Tables \ref{comp_results}(a--b), our results on $R$ and $I_{\rm Fe}$ approximately agree with the previous measurements.

\subsubsection{Close and distant reflector}
\label{close_distant}

We see in Table \ref{comp_results}(a) that the relative strength of reflection,
$R$, is close to being twice higher in the dim state than in the bright one. (In
the medium state, the amount of data is very limited, and the apparent lack of
detectable reflection may be not typical to that flux level.) The increase of
$R$ with decreasing flux could be a real change of the solid angle subtended by
the reflector between the states. On the other hand, it could be that it is due
to the contribution of a distant reflector, in particular a molecular torus, as
postulated in the AGN unified model. Reflection from distant media is indeed
commonly seen in Seyferts 2, e.g., \citet{Reynolds1994}. If we attribute the
increase of $R$ in the dim state due to that component, we can calculate the
fractional reflection strength from the close reflector, presumably an accretion
disc, and the flux reflected from the torus, which, due to its large size, is
averaged over a long time scale, and thus assumed to be the same in either
state. 

The hypothesis of the stable geometry of the source is supported by the approximate constancy of the Compton $y$ parameter, see Table \ref{comp_results}(a) and Fig.\ \ref{contours}, and of the photon index, $\Gamma$, see Table \ref{pl_fits}. Those parameters are closely related to the amplification factor of Comptonization (e.g., \citealt{Beloborodov1999}), whose constancy can be most readily achieved if the system geometry is constant (e.g., \citealt{Haardt1991,Zdziarski1999}). Thus, we consider an approximately constant relative reflection strength from the disc very likely. Observationally, we do see a strong $R$-$\Gamma$ correlation in Seyferts and X-ray binaries, which can be interpreted by a geometrical feedback model \citep{Zdziarski1999}, and which would require a constant disc $R$ for a given $\Gamma$.

The unabsorbed flux in a given state, $F_{\rm S}$, where ${\rm S}=$ either D (dim) or B (bright), is the sum of the (unabsorbed) incident (i) and reflected (r) fluxes,
\begin{equation}
F_{\rm S}=F_{\rm i,S}+F_{\rm r,S},
\label{decomposition}
\end{equation}
where $F_{\rm S}$ and $F_{\rm r,S}$ can be calculated from the results of our spectral fitting in Section \ref{Compton}. We assume that the observed (unabsorbed) reflected flux in each state is the sum of the close disc (d) state-dependent component and the constant distant torus (t) component,
\begin{equation}
F_{\rm r,S} = F_{\rm d,S} + F_{\rm t}.
\label{sum}
\end{equation}  
Then, assuming the constancy of the solid angle subtended by the disc reflector, or, equivalently, its reflection strength, $R_{\rm d}$, the disc-reflected fluxes are,
\begin{equation}
F_{\rm d,S} =F_{\rm r,S} (R_{\rm d}/ R_{\rm S}),
\label{disc_fluxes}
\end{equation}
where $R_{\rm S}$ is the reflection strength for a given state. We have four equations (\ref{sum}--\ref{disc_fluxes}) for four unknowns, $F_{\rm d,B}$, $F_{\rm d,D}$, $R_{\rm d}$ and $F_{\rm t}$, which can be readily solved. We note that they can be formulated without involving $F_{\rm i,S}$, which implies that the solution does not depend on the choice of the energy interval in which the fluxes are measured as long as it includes the entire reflected spectrum. The solutions for $R_d$ and $F_{\rm t}$ are,
\begin{equation}
R_{\rm d} = \frac{(F_{\rm r,B}-F_{\rm r,D})R_{\rm B}R_{\rm D}}{F_{\rm r,B} R_{\rm D}-F_{\rm r,D} R_{\rm B}}, \qquad
F_{\rm t} = \frac{(R_{\rm D}-R_{\rm B})F_{\rm r,B}F_{\rm r,D} } {F_{\rm r,B} R_{\rm D}-F_{\rm r,D} R_{\rm B}}. \label{solution}
\end{equation}  
Then, the fractional reflection strength of the torus can be approximately estimated as, 
\begin{equation}
R_{\rm t}\simeq  \langle R\rangle (F_{\rm t}/ \langle F_{\rm r}\rangle),
\label{torus}
\end{equation}
where $\langle R\rangle$ and $\langle F_{\rm r}\rangle$ are the average observed reflection strength and the average reflected flux, respectively.

Using the values obtained for thermal Comptonization using escape probability 
formalism (geometry parameter $=0$), we obtain $R_{\rm d} = 0.27\pm 0.07$, 
$F_{\rm t}$ = { (8.2$\pm$0.6)}$\times 10^{-11}$ erg cm$^{-2}$ s$^{-1}$, and 
$R_{\rm t}$ = { 0.24$\pm$0.04} using either arithmetic or geometric averages 
in equation (\ref{torus}). Similar values are obtained for other Comptonization 
geometries considered in Section \ref{Compton}. We can check the robustness of 
our estimates by allowing the reflection strength to depend on the accretion 
rate, or flux. If we assume that $R_{\rm d}$ in the dim state is a half of that 
in the bright state [which requires an appropriate change of equation 
(\ref{disc_fluxes})], we obtain relatively similar values of the disc reflection 
in the bright state of $R_{\rm d,B}$ = { 0.23$\pm$0.12}, $F_{\rm t}$ = 
{ (1.1$\pm$0.1)}$\times 10^{-10}$ erg cm$^{-2}$ s$^{-1}$, and $R_{\rm t}$ =
{ 0.32$\pm$0.04}. 

Similar reasoning can be applied to the Fe K$\alpha$ line. The observed line 
photon flux in the state S, $I_{\rm S}$, is assumed to be the sum of 
contributions from the disc, $I_{\rm d,S}$ and the torus, $I_{\rm t}$. The 
latter is assumed constant, and the former proportional to the ionizing flux 
incident on the disc. As a simplification, we assume the ionizing flux 
($\geq 7.1$ keV for neutral Fe) to be proportional to the differential continuum 
photon flux, $F(E)/E$, at the line centroid energy, which we denote $N_{\rm S}$, 
and which is equal to $I_{\rm S}/w_{\rm S}$, where $w_{\rm S}$ is the observed 
equivalent width in a given state. This then implies a constant disc line 
equivalent width, $w_{\rm d}$. Above, for the sake of simplicity, we have dropped 
the indices 'Fe', used in Section \ref{Compton} and Table \ref{comp_results}(b). 
The equations are, 
\begin{equation}
I_{\rm S} = I_{\rm d,S}+I_{\rm t}, \qquad
I_{\rm d,S} = w_{\rm d} N_{\rm S}, \label{line}
\end{equation}  
which can be solved for
\begin{equation}
w_{\rm d} = {I_{\rm B}-I_{\rm D}\over N_{\rm B}-N_{\rm D}}, \qquad
I_{\rm t} = {I_{\rm D} N_{\rm B}-I_{\rm B}N_{\rm D}\over N_{\rm B}-N_{\rm D}}. \label{line_solution} 
\end{equation}  
The equivalent width of the torus line flux with respect to the average incident photon flux, $ \langle N\rangle$, can be estimated as,
\begin{equation}
w_{\rm t}={I_{\rm t}/ \langle N\rangle}.
\label{wt}
\end{equation}

For the numerical values, we use here averages of the fit results for both EPIC
pn spectra in the bright state, and for the EPIC pn and XIS spectra in the dim
state, see Table \ref{comp_results}(b). We find $I_{\rm B}$ = { (2.9$\pm$0.2)}
$\times 10^{-4}$ cm$^{-2}$ s$^{-1}$, $I_{\rm D}$ = { (2.3$\pm$0.1)}$\times 
10^{-4}$ cm$^{-2}$ s$^{-1}$, $N_{\rm B}$ = { (3.6$\pm$0.1)}$\times 10^{-3}$ 
keV$^{-1}$ cm$^{-2}$ s$^{-1}$ and $N_{\rm D}$ = { (8.1$\pm$0.1)}$\times
10^{-4}$ keV$^{-1}$ cm$^{-2}$ s$^{-1}$. This yields $w_{\rm d}$ = { 23$\pm$11}
eV, $I_{\rm t}$ = { (2.1$\pm$0.9)}$\times 10^{-4}$ cm$^{-2}$ s$^{-1}$, and 
$w_{\rm t}$ = { 93$\pm$41} eV, { 120$\pm$53} eV using the arithmetic or 
geometric average, respectively.

The equivalent width of the Fe line at $i=45\degr$ by an isotropically 
illuminated cold disc for $\Gamma = 1.75$ is $\simeq 140$ eV \citep{George1991}. The value of $R_{\rm d}\simeq 0.27$ found based on the observed reflection strength then implies $w_{\rm d}\sim 38$ eV, somewhat higher than the estimate based on the line fluxes, but still in an approximate agreement taking into account measurement errors and a number of assumptions we have made. On the other hand, $R_{\rm t}\simeq 0.24$ found above would explain only $w_{\rm t}\sim 30$ eV. However, we note that we have neglected the local absorber, which obviously also gives rise to an Fe K$\alpha$ line component. Its characteristic $N_{\rm H}$ of $\sim 10^{23}$ cm$^{-2}$ (Table \ref{comp_results}b) can readily explain the excess equivalent width of $\sim 70$ eV (e.g., \citealt{Makishima1986,Awaki1991}) of the constant line component with respect to that expected from the torus. 

We note a number of caveats for our results. The standard accretion disc is flared, not flat \citep{Shakura1973}. This, however, would have a relatively minor effect, changing somehow the distribution of the inclination angles. Our results indicate disc reflection with $R_{\rm d}$ substantially less than unity, which implies the X$\gamma$ source is not entirely above the disc. A likely geometry explaining it is a hot inner flow surrounded by a disc (e.g. \citealt{Abramowicz1995,Narayan1995,Yuan2001}), in which case the incident radiation will have much larger incident angles (measured with respect to the axis of symmetry) than those assumed in the used model \citep{Magdziarz1995}. The disc may be warped (e.g., \citealt{Wijers1999}), which again would change the distribution of the incident angles. Furthermore, Compton scattering in the hot plasma above the disc reduces the observed reflection strength \citep{Petrucci2001b}.

Then, we also used the slab geometry for the torus reflection. Thus, the obtained solid angle does not correspond to the actual angle subtended by the torus from the X-ray source. For example, \citet{Murphy2009} considered a torus with a circular cross section subtending (as seen from the centre) a $2\upi$ solid angle. They found that the reflection component in this case is several times weaker than that corresponding to a slab subtending the same angle. One obvious effect here is that, unlike the case of a slab, an observer sees only a fraction of the reflecting surface, without parts obscured by the torus itself. If we take this into account, our value of $R_{\rm t}\sim 0.2$ appears consistent with a torus subtending a $\sim 2\upi$ solid angle. We note, however, that even if the torus solid angle is formally $2\upi$ or so, a large part of it will be shielded from the X-ray source by the accretion disc and the black hole itself, so the actual irradiated solid angle may be substantially lower. This effect was not taken into account in the geometrical model of \citet{Murphy2009}. Furthermore, the cross section of the torus may be substantially different from circular, which may increase the observed reflection, as well as it may be clumpy \citep*{Krolik1988,Nenkova2002}, which will decrease it. Still, our value of $w_{\rm t}\sim 100$ eV agrees with that from a torus with the column density of $N_{\rm H}\sim 10^{24}$ cm$^{-2}$ for $\Gamma\simeq 1.75$ \citep{Murphy2009}, which provides an explanation for that equivalent width alternative, or additional, to that as being due to the line emission of the absorber (discussed above). 

We have assumed that the torus reflection component is constant on the time scale of years. On the other hand, \citet{Minezaki2004} found that the dusty torus inner boundary is at $\simeq 0.04$ pc. Thus, a fraction of the torus reflection may vary on the corresponding time scale of $\sim 50$ d. Furthermore, 0.04 pc corresponds to $\simeq 2\times 10^4 R_{\rm g}$ (where $R_{\rm g}\equiv GM/c^2$ is the gravitational radius), where the accretion disc may still be present and join onto the torus. On the other hand, \citet{Radomski2003} have constrained the torus outer boundary to $\la 35$ pc, so the bulk of the torus reflection may still be constant over a time scale of years.

\subsection{Absorber properties}
\label{absorber}

We compare our results with those based on \sax, which also provided broad-band spectra, allowing to simultaneously determine absorber properties and the continuum. They were studied by \citet{Puccetti2007} and \citet{DeRosa2007}, who used the same absorber model as in this work. We find their results to be compatible with ours. The fully covering absorber has $N_{\rm H} \simeq (0.9$--$9.4)\times 10^{22}$ cm$^{-2}$ for \sax\/ and $\simeq (3.0$--$8.1)\times 10^{22}$ cm$^{-2}$ in our case, see Table \ref{comp_results}(b). For the partially covering absorber, $N_{\rm H}^{\rm p}\simeq  (3.5$--$30.3)\times 10^{22}$ cm$^{-2}$ (\sax) and $\simeq (9.5$--$30.2)\times 10^{22}$ cm$^{-2}$ (this work). The covering fractions are also similar, $f^{\rm p}\simeq 0.34$--0.71 (\sax) and 0.36--0.71 (this work). The only exception is the long-exposure \sax\/ observation of NGC 4151 in December 2001 showing a very low $N_{\rm H}$ \citep{DeRosa2007}. 

We find an anticorrelation between the $N_{\rm H}$ of both absorber 
components and the hard X-ray flux. We define the total column, $N_{\rm H}=N_{\rm 
H}^{i}+ f^{p} N_{\rm H}^{p}$. For the bright state, we have $N_{\rm H}/10^{22}$ 
cm$^{-2}$ = { 11.1$\pm$3.5} (\xmm\/ X1+X2) and { 8.2$\pm$4.9} (X3), for 
the medium state,  { 18.0$\pm$2.8} (\xmm\/ X5), and for the dim state, 
{ 23.1$\pm$0.8} (\xmm\/ X4) and { 23.2$\pm$0.9} (\suzaku). Also, the 
covering fraction is anti-correlated with the hard X-ray flux, see Table 
\ref{comp_results}(b). A similar trend is seen for the results of \citet{Puccetti2007}, who used the 6--10 keV for the X-ray flux. The variable part of the absorber needs to be relatively close to the X-ray source to be able to follow its flux on the time scale of days. The absorber in NGC 4151 was identified with massive outflows from the accretion disc \citep{Piro2005}, a broad-line region \citep{Puccetti2007} or the surface or wind of the torus \citep{Schurch2002}. Only the disc wind is close enough to the X-ray source to follow the change of the nuclear emission and to produce any correlation. The physical explanation for the correlation remains unclear; possibly the wind rate in NGC 4151 decreases with increasing accretion rate. 

\subsection{The nature of the X-ray source}
\label{xray}

The geometry and parameters of the source are constrained by a number of our 
findings. We find (i) that both the Compton parameter and the X-ray spectral 
index are approximately the same in both bright and and dim states (Sections 
\ref{cutoff}--\ref{Compton}). This implies an approximately constant 
amplification ratio of the Comptonization process, i.e., the ratio of the power 
emitted by the plasma to that supplied to it by seed photons (e.g., 
\citealt{Beloborodov1999}). The amplification factors obtained with the 
Comptonization model are indeed similar, $A$ = { 17$\pm$2, 15$\pm$2} for the 
bright and dim state, respectively (assuming $kT_{\rm bb}=10$ eV). A similar value of $A\simeq 13$  was obtained for a Comptonization model fitted to the 1991 data from \rosat, \ginga\/ and OSSE \citep{Zdziarski1996a}.

\begin{figure}
\includegraphics[width=\columnwidth]{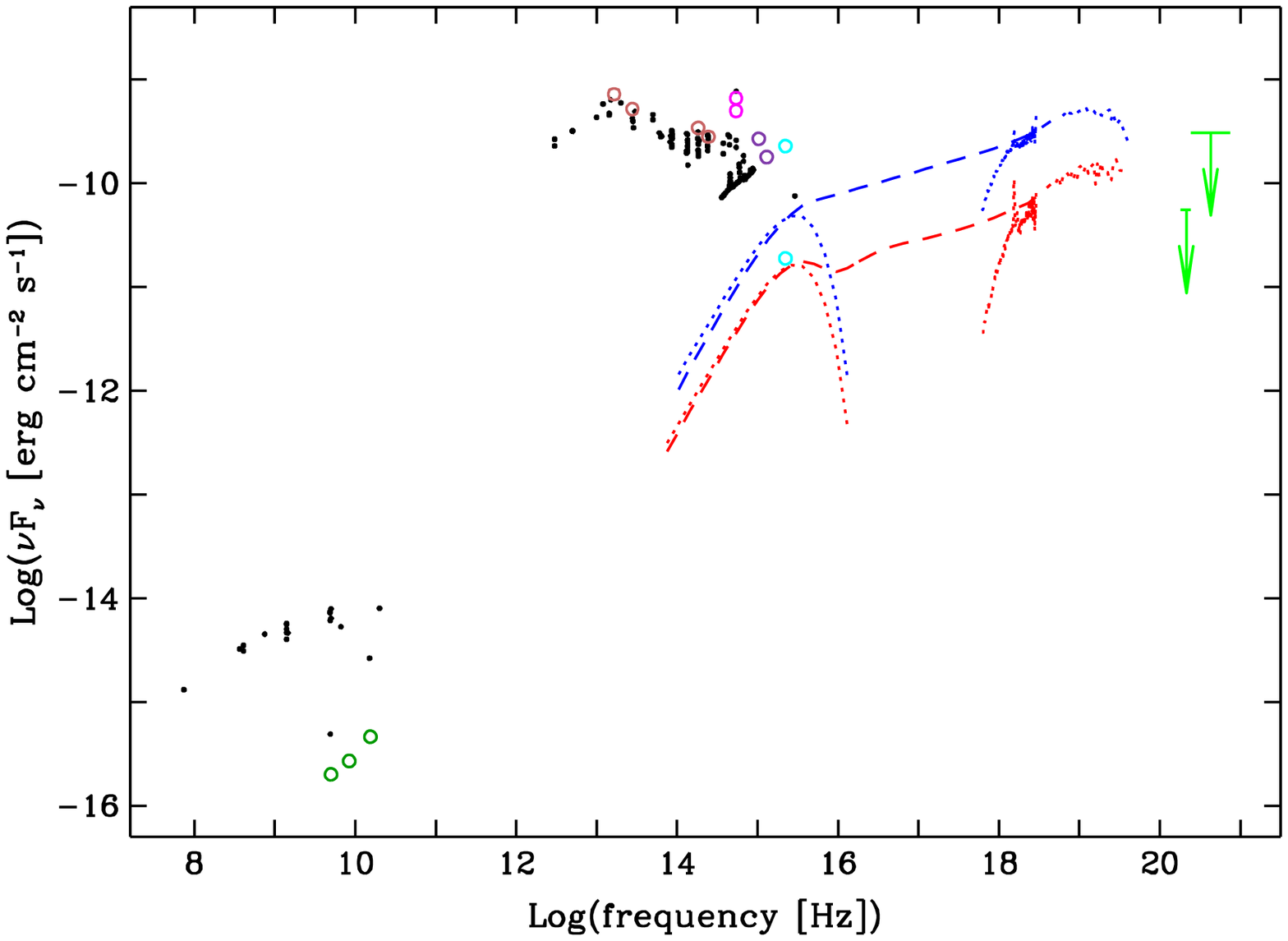}
\caption{The broad-band spectrum of NGC 4151. The radio, IR, optical and
UV data (black dots) were taken from the NED database. Additional data
are shown in circles in radio (green, \citealt{Ulvestad2005}), mid and near
IR (brown, \citealt{Radomski2003,Ruiz2003}, optical (magenta, OMC, this
work) and UV (purple, \swift/UVOT, this work; and cyan, \hst/STIS, the flux extrema from fig.\ 3 of \citealt{Kraemer2006}). The \xmm\/ EPIC pn and ISGRI spectra of the bright and dim states are shown by the blue and red dots, respectively, the corresponding unabsorbed Comptonization models below 10 keV are shown by dashes, and the disc blackbody (assuming geometry parameter $=0$ and $kT_{\rm bb}=5$ eV) incident on the hot plasma are shown by the dotted curves. The upper limits in the 0.75--1--3 MeV energy ranges are from COMPTEL \citep{Maisack1995}.}
\label{sed}
\end{figure}

If the seed photons are supplied by an accretion disc surrounding the hot
plasma, this implies an approximately constant disc inner radius. Based on this,
we have inferred (ii) a relatively weak reflection from the disc, $R_{\rm d}\simeq 0.3$ (Section \ref{close_distant}). This qualitatively agrees with $A\gg 1$ and both findings rule out the static disc corona geometry; the (outer) disc subtends a small solid angle as seen from the X-ray source, and the X-ray source subtends a small solid angle as seen from inner parts of the disc \citep{Zdziarski1999}. An implication of the latter is that the modelled disc blackbody emission (which provides seed photons for Comptonization) should be much weaker than that observed (which corresponds to the entire disc emission). This seems to be indeed the case as shown in Fig.\ \ref{sed}, where we see that the bright-state UV flux inferred from the Comptonization model is about an order of magnitude below the shown maximum observed far UV (1350 \AA; 9.2 eV) flux. On the other hand, the model dim-state far UV flux is close to the historical minimum observed. However, that minimum represented a single isolated dip in the light curve \citep{Kraemer2006}, and the actual far UV flux corresponding to the dim state is likely to be significantly higher. In choosing the shown range of the far UV flux we used its strong correlation with the optical flux, which can  be seen by comparing Fig.\ \ref{longest}(d) with fig.\ 1 in \citet{Kraemer2006}.
 
The IR, optical and UV fluxes shown in Fig. \ref{sed} appear relatively weakly affected by the host galaxy emission. In particular, the dominance of the AGN in the U band is shown by its strong variability \citep{Czerny2003}. In the optical range, there can be some non-negligible fraction of emission from the broad and narrow line regions, but in the UV we expect that the disc dominates. Given that the measured UV emission is, furthermore, absorbed by the host galaxy, the conclusion above that only a small fraction of the disc emission undergoes Comptonization appears secure. 

Another result is (iii) the connection of the normalization of the fitted disc blackbody seed emission to the disc inner radius, $R_{\rm in}$. Based on the definition from {\sc xspec}, we can express it as
\begin{equation}
K={10^8 r\cos i\over f_{\rm col}^4} {(R_{\rm in}/10^{12}\,{\rm cm})^2\over (D/10\,{\rm Mpc})^2},
\label{norm}
\end{equation}
where $f_{\rm col}\sim 1.7$ is the colour correction to the blackbody temperature \citep{Shimura1995}, and $r$ is the ratio of the blackbody emission incident on the plasma to that emitted by the disc. It implies (at $D=13.2$ Mpc and $i=45\degr$),
\begin{equation}
R_{\rm in}\simeq 4.5\times 10^{13} \left(K \over 10^{9}\right)^{1/2} \left( r\over 0.1\right)^{-1/2} \left(f_{\rm col}\over 1.7\right)^2\, {\rm cm}. 
\end{equation}
At $M=4.6\times 10^7\msun$ \citep{Bentz2006}, $R_{\rm g}\simeq 6.8\times 10^{12}$ cm. Thus, the values of $K$ in Table \ref{comp_results}(a) for the bright state (at $T_{\rm bb}=10$ eV) imply $R_{\rm in}\sim 6 R_{\rm g}$, the last stable orbit for a non-rotating black hole. This is in conflict with the findings above, according to which the inner part of the flow is occupied by the hot plasma and not by the disc, unless the black hole is rotating fast. Furthermore, the {\sc diskbb} model used for the seed photons becomes invalid when $R_{\rm in}$ is close to the last stable orbit with the standard zero-stress boundary condition \citep{Gierlinski1999}. We note, however, the disc flux is $\propto R_{\rm in}^2 T_{\rm bb}^4$, and approximately $R_{\rm in}\propto T_{\rm bb}^{-2}$ for a given X$\gamma$ spectrum. In particular, $K=7.2\times 10^9$ for the bright state at $kT_{\rm bb}=5$ eV (consistent with the UV data) and the geometry parameter $=0$, confirming the above scaling. This yields $R_{\rm in} \simeq 18 R_{\rm g}$, allowing the existence of a hot inner flow. Given that agreement, we used $kT_{\rm bb}=5$ eV in the models shown in Fig.\ \ref{sed}. We note that if $R_{\rm in}\simeq$ constant, $kT_{\rm bb}$ has to be somewhat lower in the dim state. 

For comparison, the radius, $R_{\rm eq}$, at which the integrated gravitational 
energy dissipation at $R>R_{\rm eq}$ equals that at $R<R_{\rm eq}$ is 
$R_{\rm eq}\simeq 14.7 R_{\rm g}$ for a non-rotating black hole in the 
pseudo-Newtonian approximation \citep{Paczynski1980} and with the zero-stress 
boundary condition at the last stable orbit, and less for a rotating black hole. 
Thus, $R_{\rm in}\sim R_{\rm eq}$, as found above, is consistent with the 
observed energy distribution of NGC 4151, showing the optical/UV emission 
(presumably due to the disc) at a similar level as that at $\sim 100$ keV (due 
to the hot flow). Note that $R_{\rm in}$ moderately larger than $R_{\rm eq}$ is 
required if some fraction of the energy dissipated below $R_{\rm in}$ is 
advected to the black hole instead of being radiated away. However, a disc 
truncated at $R_{\rm in}\gg R_{\rm eq}$ would be in strong conflict with the 
broad-band spectrum, as it implies the disc UV emission much weaker than that 
of X-rays. A prediction of this result is a presence of an Fe K$\alpha$ line only moderately relativistically broadened (as found earlier by Z02). 

The reflection of $R_{\rm d} \simeq 0.3$ may be easily achieved if the central 
hot region (with a large scale height) is surrounded by a disc. We have 
calculated the reflection strength in some simple geometrical models, with a 
sphere and cylinder surrounded by a flat disc. For the sphere and cylinder with 
the height equal to 1/2 of its radius we find $R_{\rm d}\simeq 0.29$ and 0.26, 
respectively, in agreement with our estimate. At this $R_{\rm d}$, the emission 
of the parts of the disc close to the hot flow is most likely dominated by 
reprocessing of X-rays, in agreement with the constancy of the $y$ parameter 
\citep{Zdziarski1999}. As a caveat, we point out that the geometrical 
$R_{\rm d}$ estimates are highly simplified, not taking into account the actual 
hot flow geometry and the radial distribution of the Comptonized emission. We 
also note that the constancy of the $y$ parameter can also be explained in a 
model in which the inner disc radius increases in the dim state, but at the 
same time the energy dissipation in the hot flow becomes relatively stronger 
in its outer region, e.g., due to advection. 

On the other hand, the effective solid angle subtended by the hot flow as seen 
from the disc can be identified with our parameter $r\sim 0.1$ above. All the 
above considerations very strongly argue for the Comptonizing plasma in NGC 4151 
being photon starved. This is not compatible with a static corona above the disc 
\citep{Haardt1991}, and compatible with a hot inner flow surrounded by an outer 
cold disc. However, it is also possible that the corona does cover the disc but 
it is in a state of a mildly relativistic outflow \citep*{Beloborodov1999,Malzac2001}. The relativistic motion then both reduces 
the flux of the up-scattered photons incident on the disc, which explains the 
relative weakness of the disc reflection, and reduces the flux of the disc 
blackbody photons entering the corona in its comoving frame, which explains 
the photon starvation of the hot plasma, i.e., $A\gg 1$. A prediction of this 
model is the presence of a very broad relativistic Fe line (as the disc now 
extends all the way to the minimum stable orbit). The lack of detection of 
such a line in high-quality \xmm\/ and \suzaku\/ spectra represents an argument
against this model.

Our next finding has been that (iv) the plasma temperature increased by a factor of $\simeq 3.5$ when the bolometric X-ray luminosity decreased by about the same factor, see Table \ref{comp_results}(a). In agreement with the approximate constancy of $y$, the optical depth also decreased by a similar factor. This can be explained if the accretion rate, $\dot M$, is $\propto L_{\rm X\gamma}$ and the hot flow has a velocity profile independent of $\dot M$, so $\tau\propto L_{\rm X\gamma}$. The temperature then follows from the energy balance provided the ratio of the power in the hot flow emission to that in seed photons cooling remains constant. However, models of accreting hot flows yield a dependence of $\tau$ on $L_{\rm X\gamma}$ weaker than the simple proportionality [e.g., eqs.\ (10--11) in \citealt{Z98}; table 2 in \citealt{Yuan2004}]. 

The total model luminosity of the Comptonization component (including reflection) in the bright and dim state is $L_{\rm X\gamma}\simeq 5.2 \times 10^{43}$ erg s$^{-1}$ and $1.5\times 10^{43}$ erg s$^{-1}$, respectively (see Table \ref{comp_results}a). The estimated average luminosity from the IR to UV is $\simeq 7 \times 10^{43}$ erg s$^{-1}$, with the IR band ($<10^{14}$ Hz) contributing $\sim 60$ per cent of it. If we then assume that the IR flux remains constant and the optical--UV flux scales as the high-energy emission, the total bolometric luminosity can be estimated as $1.3\times 10^{44}$ erg s$^{-1}$, $7\times 10^{43}$ erg s$^{-1}$ in the bright and dim state, respectively. Assuming the hydrogen mass fraction of $X=0.7$ and $M=4.6 \times 10^{7} \msun$, the Eddington luminosity, $L_{\rm E}\equiv 8\upi GM m_{\rm p} c/[(1+X)\sigma_{\rm T}]\simeq 6.8 \times 10^{45}$ erg s$^{-1}$, giving the Eddington ratios of $\simeq 0.01$--0.02. Note the factor of 2 variability of the bolometric luminosity compared to the factor of 3.5 variability of the Comptonization component. This difference appears to be explained by the presence of the constant, averaged over very long time scales, IR component. 

Relevant time scales include the light crossing time across the circumference of the last stable orbit, which for a non-rotating black hole is $\sim 2700$ s (at the default mass). The corresponding light crossing time at $R_{\rm in}\sim 15 R_{\rm g}$ is $\sim 7000$ s. Thus, the hour-time scale variability observed by ISGRI in the bright state (Fig.\ \ref{short}) appears to correspond to fluctuations at those radii. On the other hand, the viscous time of an accretion flow is,
\begin{eqnarray}
\lefteqn{t_\mathrm{visc}= {GM\over c^3} \left(R\over R_{\rm g}\right)^{3/2} \alpha^{-1} \left(H\over R\right)^{-2} }\\
\lefteqn{
\qquad\!\! \simeq 1.2\times 10^6\,{\rm s} \left(R\over 15R_{\rm g}\right)^{3/2}
\left(M\over 4.6\times 10^7\msun\right) \left(\alpha\over 0.1\right)^{-1} \left(H/R\over 1/3\right)^{-2},\nonumber}
\label{eq:tvisc}
\end{eqnarray}
where $H$ is the flow scale height, $\alpha$ is the viscosity parameter, and $H/R\sim 1/3$ for a hot flow. At our default mass, $15 R_{\rm g}$ and $\alpha=0.1$, $t_\mathrm{visc}\simeq 13$ d. For comparison, the \swift/BAT data for NGC 4151 need to be rebinned to several days in order to get a sufficient signal-to-noise ratio for low flux states. Thus, this detector can sample the overall behaviour of the hot flow but not variability close to the last stable orbit. 

The spectrum of NGC 4151 resembles spectra of black-hole binaries in the hard state, also commonly fitted by thermal Comptonization. However, a faint non-thermal tail has occasionally been seen there, e.g., \citet{McConnell2002}, \citet{Wardzinski2002}. The Comptonization model used by us also allows to include a corresponding non-thermal tail to the electron distribution. However, the data for NGC 4151 do not allow us to test for the presence of any non-thermal spectral component, see Section \ref{Compton}. 

\subsection{Comparison with other objects}
\label{other}

We have presented the average parameters of Seyfert spectra in Section \ref{intro}, for both e-folded power law and Comptonization models. Our results for NGC 4151 are within the range observed for other Seyferts, as noted by Z02. Z02 also provide a critical analysis of earlier finding of very hard X-ray spectra in NGC 4151, which they find to result from improper modelling of its complex absorption. NGC 4151 in the bright state is only slightly harder than the average Seyfert 1, and it shows less reflection, in accordance with the reflection-index correlation \citep{Zdziarski1999}. On the other hand, the dim state appears to have a significant contribution to reflection from a distant reflector, possibly a torus (Section \ref{close_distant}). 

The similarity between spectra of Seyferts and black-hole binaries in the hard spectral state was pointed out by, e.g., \citet{Zdziarski1996b}, though the latter appear to have on average somewhat harder spectra. This is likely to be due to the disc seed photon energies being much higher than those in Seyferts, which reduces the spectral index, $\Gamma$, for a given Comptonization amplification factor, $A$ \citep{Beloborodov1999}. 

A striking similarity between the OSSE spectra of the black-hole binary GX 339--4 in the hard state and NGC 4151 was pointed out by \citet{Zdziarski1998}. \citet{Wardzinski2002} have shown that when $L_{\rm X\gamma}$ of GX 339--4 decreased by a factor of $\sim 2$, $kT_{\rm e}$ increased significantly (see their fig.\ 1a), from $46^{+6}_{-4}$ keV to $76^{+5}_{-6}$ keV, with only a slight softening of the X-ray slope. This effect is very similar to that seen in NGC 4151, see Table \ref{comp_results}(a). In both NGC 4151 and GX 339--4, this effect may be due to a decrease of $\dot M$ in the hot flow causing the associated reduction of $\tau$.

\section{Conclusions}
\label{conclusions}

We have presented a comprehensive spectral analysis of all \integral\/ data obtained so far for NGC 4151, together with all contemporaneous data from \xte, \xmm, \swift\/ and \suzaku\/ (Section \ref{data}). Our main findings are summarized below.

We have found that the 20--100 keV emission measured by \integral\/ has had almost the same range of fluxes as that measured by other satellites during past 40 years (Section \ref{variability}). Thus, our analysis appears to explore the full range of the variability of this object. Also, we have found that this flux appears to uniquely determine the intrinsic broad-band spectrum. 

Simultaneous observations by \integral\/ in the optical range and in medium and hard X-rays show a very strong correlation within the X-ray band, and a less clear correlation between the optical and X-ray emission, with a bimodal behaviour of the optical flux. The linearity of the former correlation shows that the X-ray spectral slope of NGC 4151 remains almost constant.

Most of the \integral\/ observations correspond to either a bright or dim hard X-ray state (Section \ref{assumptions}). We have found that a thermal Comptonization model provides very good fits to the data (Section \ref{Compton}). As the state changes from bright to dim (with $L_{\rm X\gamma}$ decreasing by a factor of $\simeq 3.5$), the plasma electron temperature increases from $kT_{\rm e}\sim 60$ keV to $\sim 200$ keV and the Thomson optical depth decreases from $\tau\sim 2$ to $\sim 0.6$ (for spherical geometry), i.e., $\tau\propto 1/kT_{\rm e}$ and $\tau \propto L_{\rm X\gamma}$. The former proportionality corresponds to an almost constant Compton $y$ parameter, with the X-ray slope remaining almost constant, with only a slight softening in the dim state, in agreement with the medium/hard X-ray correlation. This is suggestive of almost constant source geometry and amplification factor. The latter proportionality may occur due to the accretion rate varying at an approximately constant hot flow velocity. 

The fitted strength of Compton reflection increased from $R\simeq 0.4$ to $R\simeq 0.8$, which, at the face value, would be in conflict with the constant source geometry indicated by the constant $y$. However, in accordance with the AGN unified model, NGC 4151 is likely to possess a remote torus, also reflecting the central X-ray emission. Assuming the solid angles subtended by both the (close) disc reflector and the (distant) torus reflector are constant, we can explain the varying fitted reflection strength by the presence of a contribution of the flux reflected by the torus, which is constant over the observation time scale given its very large size (Section \ref{close_distant}). In the bright state, the contribution of the torus reflection is small, but it becomes much stronger in the dim state, explaining the fitted large value of $R$. Given the above assumption, we find that the solid angles subtended by the disc and torus (as seen from the X-ray source) are $\simeq 0.3$, $0.2$ of $2\upi$, respectively. These values have been obtained with a slab reflection model, which is likely to substantially underestimate the actual solid angle subtended by the torus, which might then be close to $2\upi$. 

The Comptonizing plasma is photon-starved, i.e., the flux in seed photons incident on the plasma is $\sim$15 times weaker than the Comptonized flux (Section \ref{xray}). This is consistent with the seed photons being supplied by a truncated optically-thick disc surrounding a hot accretion flow. This geometry is also consistent with the relatively small disc reflection. We also find that the flux of the disc photons fitted to the X-ray spectra is an order of magnitude below the UV fluxes actually observed. This is again consistent with the above geometry, in which only a small fraction of the emitted disc photons is incident on the inner hot regions. All these findings rule out a static disc corona geometry. However, they are still compatible with a corona outflowing at a mildly relativistic speed, see Section \ref{xray}. 

In any case, the disc inner radius cannot be very large given that the observed $EF_E$ at UV are of the same order as those at hard X-rays, implying a rough equipartition between the disc and hot flow emission. For a non-rotating black hole, the radius with the same integrated dissipation below and above it is $\simeq 15 R_{\rm g}$. The truncation radius can also be constrained by the fitted Comptonization model, which assumes the seed photons are disc blackbody. This gives a similar value of $R_{\rm in}$ for the inner disc temperature of $\sim 5$ eV, which value is consistent with the observations. 

Compared to Seyfert 1s, NGC 4151 has a somewhat harder X-ray spectrum and less reflection. Its spectrum is also similar to black-hole binaries in the hard state, in particular to GX 339--4, which has shown a very similar spectral evolution. 

Future missions, e.g., \textit{Astro-H\/} and \textit{EXIST}, with 
instruments with a sensitivity comparable to that of OSSE, can provide better 
data above 200 keV than obtained up to date. Yet, for a better understanding 
of the physics of the central engine of AGNs, we need high-quality data up to 
at least 1 MeV. This would yield strong constraints on the electron distribution 
in the Comptonization region (e.g. the presence of a non-thermal tail) and on 
the source geometry (e.g. the spatial distribution of the electron temperature 
and the characteristic optical depth.

\section*{Acknowledgments}
We thank B. Czerny for valuable discussions. PL and AAZ have been supported in part by the Polish MNiSW grants NN203065933 and 362/1/N-INTEGRAL/2008/09/0. We used data from the High Energy Astrophysics Science Archive Research Center, and from the NASA/IPAC Extragalactic Database. 

\bibliographystyle{mnras}
\bibliography{plnew}

\label{lastpage}

\end{document}